\documentclass[12pt]{article} 
\usepackage[top=30mm, bottom=40mm, left=25mm, right=25mm]{geometry}
\usepackage{cite} 

\usepackage{graphicx}
\usepackage{amsmath, amssymb} 				
\usepackage{empheq}
\usepackage{fancybox}
\usepackage{mathrsfs} 
\usepackage{textcomp} 
\usepackage{mathtools}

\usepackage{epstopdf}
\usepackage{bm} 
\usepackage{cases} 

\usepackage{latexsym}
\usepackage{marvosym}
\usepackage{bm}	

\usepackage[mathcal]{euscript}
\usepackage{indentfirst}

\usepackage[rgb]{xcolor}

\usepackage{hyperref}
\hypersetup{
	unicode,	
	colorlinks,
	citecolor=[rgb]{0,0,1},
	linkcolor=[rgb]{.5, 0, .5},
	urlcolor=[rgb]{.6,0,0},  
	bookmarksopen=true,
	bookmarksopenlevel=\maxdimen,
	bookmarksnumbered
	}

\numberwithin{equation}{section}

\usepackage[all]{xy}
\usepackage{tikz}
	\usetikzlibrary{decorations.markings}
	\usetikzlibrary{arrows}

\usepackage{amsthm} 
\theoremstyle{definition}

\theoremstyle{plain}

\font\biggest=cmssbx10 scaled 3200
\font\bigger=cmssbx10 scaled 2560

\def\K3{\mathrm K3}

\def\gl#1#2{$\mathrm{GL}(#1; {\bf #2})$}
\def\sl#1#2{$\mathrm{SL}(#1; {\bf #2})$}
\def\sp#1#2{$\mathrm{Sp}(#1; {\bf #2})$}
\def\spin#1#2{$\mathrm{Spin}(#1, #2)$}
\def\su#1#2{SU({#1,#2})}
\def\usp#1#2{USp({#1,#2})}
\def\U#1{U({#1})}

\def\O#1{O({#1})}

\def\double #1{#1{\hbox{\kern-2pt $#1$}}}
\def\un#1{\underline #1}

\def\gl#1#2{\ifmmode \mathrm{GL}(#1; {\bf #2}) \else $\mathrm{GL}(#1; {\bf #2})$\fi}
\def\sl#1#2{\ifmmode \mathrm{SL}(#1; {\bf #2}) \else $\mathrm{SL}(#1; {\bf #2})$\fi}
\def\so#1{\ifmmode \mathrm{SO}({#1}) \else $\mathrm{SO}(#1)$\fi}

\def\sp#1#2{\ifmmode \mathrm{Sp}(#1; {\bf #2}) \else $\mathrm{Sp}(#1; {\bf #2})$\fi}
\def\usp#1#2{\ifmmode \mathrm{USp}(#1,#2) \else $\mathrm{USp}(#1,#2)$\fi}
\def\spin#1#2{\ifmmode \mathrm{Spin}(#1,#2) \else $\mathrm{Spin}(#1,#2)$\fi} 
\def\su#1{\ifmmode \mathrm{SU}({#1}) \else $\mathrm{SU}(#1)$\fi}

\def\on#1#2{{\buildrel{{\mkern2.5mu\raise-.1em\hbox{$\scriptstyle#1$}\mkern-2.5mu}}\over{#2}}}		
\def\ron#1#2{{\buildrel{{\raise-.1em\hbox{$\scriptstyle#1$}}}\over{#2}}}		
\def\dt#1{\on{\hbox{\bf .}}{#1}}                
\mathcode`\*="702A                  
\def\f#1#2{{\textstyle{#1\over#2}}}	   
\def\half{{\textstyle{1\over{\raise.1ex\hbox{$\scriptstyle{2}$}}}}}
\def\slap#1#2{\setbox0=\hbox{$#1{#2}$}#2\kern-\wd0{\hbox to\wd0{\hfil$#1{/}$\hfil}}}
\def\sla#1{\mathpalette\slap{#1}}		

\def\Gamma{\mathchar"0100}
\def\Delta{\mathchar"0101}
\def\Theta{\mathchar"0102}
\def\Lambda{\mathchar"0103}
\def\Xi{\mathchar"0104}
\def\Pi{\mathchar"0105}
\def\Sigma{\mathchar"0106}
\def\Upsilon{\mathchar"0107}
\def\Phi{\mathchar"0108}
\def\Psi{\mathchar"0109}
\def\Omega{\mathchar"010A}

\catcode128=13 \def €{\"A}                 
\catcode129=13 \def {\AA}                 
\catcode130=13 \def '{\c}           	   
\catcode131=13 \def ƒ{\'E}                   
\catcode132=13 \def "{\~N}                   
\catcode133=13 \def …{\"O}                 
\catcode134=13 \def †{\"U}                  
\catcode135=13 \def ‡{\'a}                  
\catcode136=13 \def ˆ{\`a}                   
\catcode137=13 \def ‰{\^a}                 
\catcode138=13 \def Š{\"a}                 
\catcode139=13 \def ‹{\~a}                   
\catcode140=13 \def Œ{\alpha}            
\catcode141=13 \def {\chi}                
\catcode142=13 \def Ž{\'e}                   
\catcode143=13 \def {\`e}                    
\catcode144=13 \def {\^e}                  
\catcode145=13 \def '{\"e}                
\catcode146=13 \def '{\'\i}                 
\catcode147=13 \def "{\`\i}                  
\catcode148=13 \def "{\^\i}                
\catcode149=13 \def •{\"\i}                
\catcode150=13 \def –{\~n}                  
\catcode151=13 \def —{\'o}                 
\catcode152=13 \def ˜{\`o}                  
\catcode153=13 \def ™{\^o}                
\catcode154=13 \def š{\"o}                 
\catcode155=13 \def ›{\~o}                  
\catcode156=13 \def œ{\'u}                  
\catcode157=13 \def {\`u}                  
\catcode158=13 \def ž{\^u}                
\catcode159=13 \def Ÿ{\"u}                
\catcode160=13 \def  {\tau}               
\catcode162=13 \def ¢{\oplus}           
\catcode163=13 \def £{\relax\ifmmode\expandafter\to\else\expandafter\itemize\fi} 
\catcode164=13 \def ¤{\subset}	  
\catcode165=13 \def ¥{\infty}           
\catcode166=13 \def ¦{\mp}                
\catcode167=13 \def §{\sigma}           
\catcode168=13 \def ¨{\rho}               
\catcode169=13 \def ©{\gamma}         
\catcode170=13 \def ª{\leftrightarrow} 
\catcode171=13 \def «{\relax\ifmmode\expandafter\acute\else\expandafter\'\fi}
\catcode172=13 \def ¬{\relax\ifmmode\expandafter\ddt\else\expandafter\"\fi}
\catcode173=13 \def ­{\equiv}            
\catcode174=13 \def ®{\approx}          
\catcode175=13 \def ¯{\Omega}          
\catcode176=13 \def °{\otimes}          
\catcode177=13 \def ±{\ne}                 
\catcode178=13 \def ²{\le}                   
\catcode179=13 \def ³{\ge}                  
\catcode180=13 \def ´{\upsilon}          
\catcode181=13 \def µ{\mu}                
\catcode182=13 \def ¶{\delta}             
\catcode183=13 \def ·{\epsilon}          
\catcode184=13 \def ¸{\Pi}                  
\catcode185=13 \def ¹{\pi}                  
\catcode186=13 \def º{\beta}               
\catcode187=13 \def »{\partial}           
\catcode188=13 \def ¼{\nobreak\ }       
\catcode189=13 \def ½{\zeta}               
\catcode190=13 \def ¾{\sim}                 
\catcode191=13 \def ¿{\omega}           
\catcode192=13 \def À{\dt}                     
\catcode193=13 \def Á{\gets}                
\catcode194=13 \def Â{\lambda}           
\catcode195=13 \def Ã{\nu}                   
\catcode196=13 \def Ä{\phi}                  
\catcode197=13 \def Å{\xi}                     
\catcode198=13 \def Æ{\psi}                  
\catcode199=13 \def Ç{\int}                    
\catcode200=13 \def È{\oint}                 
\catcode201=13 \def É{\relax\ifmmode\expandafter\cdot\else\vol\fi}    
\catcode202=13 \def Ê{\relax\ifmmode\expandafter\,\else\thinspace\fi}
\catcode203=13 \def Ë{\`A}                      
\catcode204=13 \def Ì{\~A}                      
\catcode205=13 \def Í{\~O}                      
\catcode206=13 \def Î{\Theta}              
\catcode207=13 \def Ï{\theta}               
\catcode208=13 \def Ð{\relax\ifmmode\expandafter\bar\else\expandafter\=\fi}
\catcode209=13 \def Ñ{\overline}             
\catcode210=13 \def Ò{\langle}               
\catcode211=13 \def Ó{\relax\ifmmode\expandafter\{\else\ital\fi}      
\catcode212=13 \def Ô{\rangle}               
\catcode213=13 \def Õ{\}}                        
\catcode214=13 \def Ö{\sla}                      
\catcode215=13 \def ×{\relax\ifmmode\expandafter\check\else\expandafter\v\fi}
\catcode216=13 \def Ø{\"y}                     
\catcode217=13 \def Ù{\"Y}  		    
\catcode218=13 \def Ú{\Leftarrow}       
\catcode219=13 \def Û{\Leftrightarrow}       
\catcode220=13 \def Ü{\relax\ifmmode\expandafter\Rightarrow\else\sect\fi}
\catcode221=13 \def Ý{\sum}                  
\catcode222=13 \def Þ{\prod}                 
\catcode223=13 \def ß{\widehat}              
\catcode224=13 \def à{\pm}                     
\catcode225=13 \def á{\nabla}                
\catcode226=13 \def â{\quad}                 
\catcode227=13 \def ã{\in}               	
\catcode228=13 \def ä{\star}      	      
\catcode229=13 \def å{\sqrt}                   
\catcode230=13 \def æ{\^E}			
\catcode231=13 \def ç{\Upsilon}              
\catcode232=13 \def è{\"E}    	   	 
\catcode233=13 \def é{\`E}               	  
\catcode234=13 \def ê{\Sigma}                
\catcode235=13 \def ë{\Delta}                 
\catcode236=13 \def ì{\Phi}                     
\catcode237=13 \def í{\`I}        		   
\catcode238=13 \def î{\iota}        	     
\catcode239=13 \def ï{\Psi}                     
\catcode240=13 \def ð{\times}                  
\catcode241=13 \def ñ{\Lambda}             
\catcode242=13 \def ò{\cdots}                
\catcode243=13 \def ó{\^U}			
\catcode244=13 \def ô{\`U}    	              
\catcode245=13 \def õ{\bo}                       
\catcode246=13 \def ö{\relax\ifmmode\expandafter\hat\else\expandafter\^\fi}
\catcode247=13 \def÷{\relax\ifmmode\expandafter\tilde\else\expandafter\~\fi}
\catcode248=13 \def ø{\ll}                         
\catcode249=13 \def ù{\gg}                       
\catcode250=13 \def ú{\eta}                      
\catcode251=13 \def û{\kappa}                  
\catcode252=13 \def ü{\half}     		 
\catcode253=13 \def ý{\Gamma} 		
\catcode254=13 \def þ{\Xi}   			
\catcode255=13 \def ÿ{\relax\ifmmode\expandafter{}^{\dagger}{}\else\dag\fi}

\def\A{{\cal A}}  \def\B{{\cal B}}  \def\C{{\cal C}}  \def\D{{\cal D}}
\def\E{{\cal E}}         
  \def\K{{\cal K}}    \def\M{{\cal M}}   \def\N{{\cal N}}  
\def\O{{\cal O}}  \def\P{{\cal P}}  \def\Q{{\cal Q}}  
\def\S{{\cal S}}    \def\U{{\cal U}} \def\V{{\cal V}}  \def\W{{\cal W}}

\font\textscr=stix-mathscr at 12pt \font\scriptscr=stix-mathscr at 8pt \font\sscriptscr=stix-mathscr at 6pt
\def\scr{\fam15 \textscr}
\def\a{{\scr a}}	\def\b{{\scr b}}	\def\c{{\scr c}}	\def\d{{\scr d}}	
	\def\ff{{\scr f}}		
\def\m{{\scr m}}	 	\def\p{{\scr p}}
\def\r{{\scr r}}	\def\s{{\scr s}}

\def\^{\wedge}

\def\dd{\hbox{\,\Large$\triangleright$}}

\def\dig#1{\setbox0=\hbox{$#1M$}
	\hskip.06\wd0 \vrule width.07\wd0 height.63\wd0 depth.01\wd0 
	\vrule width.37\wd0 height.63\wd0 depth-.56\wd0 \hskip-.4\wd0
	\vrule width.25\wd0 height.35\wd0 depth-.28\wd0 
	\vrule width.07\wd0 height.35\wd0 depth-.17\wd0 \hskip.14\wd0}
\def\digamma{{\mathpalette\dig{}}}

\def\hang{\hangindent\parindent}
\def\bitem{\par\hang\textindent}
\def\textindent#1{\indent\llap{#1\enspace}\ignorespaces}

\title{
{\color{purple}\biggest F-brane Superspace:\\[0in]}
{\color{violet}\bigger The New World Volume}
}
\author{William D. Linch \textsc{iii}$\,{}^\text{\Pisces}$ and Warren Siegel$\,{}^\text{\Scorpio}$}
\date{}

\begin{document}
\maketitle

\vspace*{-8cm}
\begin{flushright}
{~\\
MI-TH-1765\\
YITP-SB-17-32}
\end{flushright}
\vspace*{+7cm}

\begin{center}
{\em
${}^{\mbox{\footnotesize\Pisces}}$
Mitchell Institute for
Fundamental Physics and Astronomy, \\
Texas A\&{}M University,
College Station, TX 77843\\
~\\
${}^{\mbox{\footnotesize\Scorpio}}$
C. N. Yang Institute for Theoretical Physics\\
State University of New York, Stony Brook, NY 11794-3840
}
\end{center} 

\vspace{10pt}

\begin{abstract}
\noindent 
F-theory requires a new Virasoro algebra, including $\kappa$-symmetry, with a worldvolume coordinate for each generator. (Similar is implied for the superstring.) Doubles of the spacetime coordinates are eliminated by selfduality, which now applies to all currents.
\end{abstract}

\vfill
\begin{flushleft}
~\\
{${}^{\mbox{\footnotesize\Pisces}}$ \href{mailto:wdlinch3@gmail.com}{wdlinch3@gmail.com}}\\
{$^{\text{\Scorpio}}$ \href{mailto:siegel@insti.physics.sunysb.edu}{siegel@insti.physics.sunysb.edu}}
\end{flushleft}

\setcounter{page}1
\thispagestyle{empty}
\newpage

{
\linespread{1.1}
\small
\tableofcontents
}

\newpage
\section{Introduction}
\label{S:Introduction}


\subsection{Preface}

F-theory (as a theory, as opposed to a set of classical vacua) is a symmetry-enhancing generalization subordinating both M(embrane)-theory and T(-duality)-theory \cite{Siegel:1993th,Siegel:1993bj}, themselves different generalizations of S(tring)-theory.  T-duality has been proven directly on the first-quantized string action; but S-duality is nonperturbative in the string coupling, and so would require a second-quantized action (superstring field theory) for its definition, and so far has been demonstrated only on the massless (supergravity) sector (and its classical solutions).  In a complete F-theory STU-duality would be at least as manifest as T-duality in T-theory:  In particular the generalization of winding modes would be included.

The cosets representing the massless, bosonic fields of the (minimal) theories
are succinctly represented in the diagram \cite{Polacek:2014cva}:
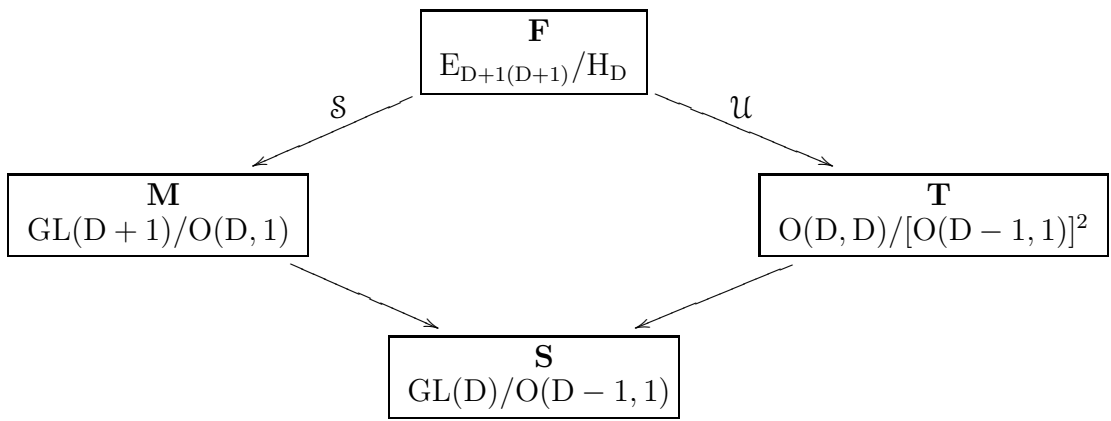
\begin{align}
\xymatrix{
	& \fbox{ \txt{ {\bf F} \\ $\rm E_{D+1(D+1)}/H_D$} \ar[dl]_{\textstyle\S} \ar[dr]^{\textstyle\U} }&\cr
\fbox{ \txt{ {\bf M}\\ $\rm GL(D+1)/O(D,1)$} \ar[dr] }&
		& \fbox{ \txt{ {\bf T}\\ ${\rm O(D,D)/[O(D-1,1)]}{}^2$} \ar[dl] } \cr
	&\fbox{ \txt{ {\bf S} \\ $\rm GL(D)/O(D-1,1)$}}&
}
\nonumber
\end{align}
The isotropy groups are symmetry groups of the current algebras (and the vacuum); they transform the fermions (in particular, supersymmetry).  
For S-theory they are simply the Lorentz group, and close relatives for M and T-theory; for F-theory they are various generalizations.
These generalized Lorentz groups follow from the general properties of spinors in dimension D mod 8 \cite{Linch:2016ipx}:
\vskip-.2in
{ 
\font\qc=manfnt scaled\magstep2

\catcode`\ =9 \endlinechar=-1 
\newcount\dir \newdimen\yy \newdimen\ww
\newif\ifvisible \let\b=\visibletrue \let\w=\visiblefalse
\newbox\NE \newbox\NW \newbox\SE \newbox\SW \newbox\NS \newbox\EW
\setbox\SW=\hbox{\qc a} \setbox\NW=\hbox{\qc b}
\setbox\NE=\hbox{\qc c} \setbox\SE=\hbox{\qc d}
\ww=\wd\SW \dimen0=\fontdimen8\qc
\setbox\EW=\hbox{\kern-\dp\SW \vrule height\dimen0 width\wd\SW} \wd\EW=\ww
\setbox\NS=\hbox{\vrule height\ht\SW depth\dp\SW width\dimen0}  \wd\NS=\ww
\def\l{\ifcase\dir \dy+\NW \or\dx-\SW \or\dy-\SE \or\dx+\NE\dd-4\fi \dd+1}
\def\s{\ifcase\dir \dx+\EW \or \dy+\NS \or \dx-\EW \or \dy-\NS \fi}
\def\r{\ifcase\dir \dy-\SW\dd+4 \or\dx+\SE \or\dy+\NE \or\dx-\NW\fi \dd-1}
\def\t{\ifcase\dir\kern-\ww\dd+2\or\ey-\dd+2\or\kern\ww\dd-2\or\ey+\dd-2\fi}
\edef\dd#1#2{\global\advance\dir#1#2\space}
\def\dx#1#2{\ifvisible\raise\yy\copy#2 \if#1-\kern-2\ww\fi\else\kern#1\ww\fi}
\def\dy#1#2{\ifvisible\raise\yy\copy#2 \kern-\ww \fi \global\advance\yy#1\ww}
\def\ey#1{\global\advance\yy#1\ww}
\def\path#1{\hbox{\b \dir=0 \yy=0pt #1}}
\catcode`\ =10 \endlinechar='15 

\begin{align*}
\kern-7em
\begin{matrix} \hfill \rm H & \rm SO & \rm SL & \rm Sp & & \textrm{H$'$ (SU)} \cr
		\noalign{\vskip.5em}
		\hfill\textrm{Real} & 1 & 2 & 3 & X^{(μ)} & \rm SO \cr
		\noalign{\vskip.5em}
		\hfill\textrm{Complex} & 0 & & 4 & X^{ŒÀº} & \rm SU \cr
		\noalign{\vskip.5em}
		\hfill\textrm{Pseudo-real} & 7 & 6 & 5 & X^{[μ]}&  \rm USp \cr
		\noalign{\vskip.5em}
		& §^{(Œº)} & §^Œ{}_{к} & §^{[Œº]} & & \cr
		& Y^{(Œ'º')} & Y^{Œ'}{}_{к'} & Y^{[Œ'º']} & & \cr
		\end{matrix}
\kern-15em\raise2.5em\path{\s\s\s\l\l\s\s\s\l\l}
\kern6.7em\raise.2em\path{\l\s\s\l\l\s\s\l}
\kern-3.4em\raise-1.8em\path{\s\s\s\l\l\s\s\s\l\l}
\kern-3.2em\raise-1.9em\path{\l\s\s\l\l\s\s\l}\kern9em
\end{align*}

} 
The (spacetime) spinor size increases by a factor of 2 from one oval to the next.  We have indicated not only the worldvolume coordinates $§$ and spacetime coordinates $X$, but also  the R-symmetry group H$'$ for worldvolume scalars $Y$, and the bispinor index structure for all.
For F-theory and S-theory, the isotropy groups are thus:
\begin{equation}
\label{E:table}
\begin{array}{rccccccc}
\rm D & 1 & 2 & 3 & 4 & 5 & 6 & 7\\
\rm H_D & \rm SO(1,1) & \rm GL(2) & \rm Sp(4) & \rm Sp(4; {\bf C} ) & \rm USp(4,4) & \rm SU*(8) & \rm SO*(16) \\
\rm SO(D-1,1) & \rm I & \rm GL(1) & \rm Sp(2) & \rm Sp(2; {\bf C} ) & \rm USp(2,2) & \rm SU*(4) & \rm SO(6,1)
\end{array}
\end{equation}
The doubling is characteristic of the unification of left and right spinors of Type II.  (Note SO*(8) = SO(6,2).)

Just as T-theory doubles the dimensions to incorporate winding number before compactification, F-theory further increases the number of dimensions to manifest U-duality.
The symmetry is spontaneously broken, and the dimensions reduced, by solving the section conditions in going to a ``unitary" gauge, reproducing the usual string formalism.  However, it might be possible to define ``covariant" gauges where the full symmetry remains manifest by including appropriate ghost degrees of freedom \cite{Siegel:2016dek}.

Previously we defined F-theory generalizations of strings with exceptional gravity in various dimensions D, described by selfdual coordinates $X(§)$ \cite{Linch:2015fya, Linch:2015qva}.  
This leads to a doubling of the usual string coordinates \cite{Duff:1989tf,Tseytlin:1990nb,Tseytlin:1990va} when reduced to T-theory.
We then generalized to critical superstrings by adding $10-$D spacetime scalar coordinates $Y$, their worldvolume duals $\widetilde Y$ (needed for constraints/section conditions), and spinor coordinates $Î$ for supersymmetry \cite{Linch:2015fca}.  
In this paper we will also double the fermionic coordinates \cite{Green:1989nn} as a natural extension:
Fermionic selfduality is then the equivalent of fermionic dimensional reduction \cite{Siegel:1994xr}.
(We'll omit $Y$, which will return in future papers.)

As we have seen repeatedly, target space and worldvolume coordinates/symmetries are mixed up in F-theory.
The worldvolume frame $\E$ and target frame $E$ are related by the orthogonality constraint.
The naive extension of the target space symmetries without analogously extending the worldvolume leads to a constraint that is incompatible with existence of the target gravitino. 
We thus will find that extension of the spacetime to include fermionic coordinates (superspace) will require the same for the worldvolume.
This differs from our previous treatments, where a nontrivial flat (``vacuum") worldvolume vielbein was required only when ``Lorentz" (H$_{\rm D}$) coordinates and current were introduced \cite{Ju:2016hla}.
But just as the usual worldvolume coordinates realize the Virasoro-like generators, their superpartners realize $û$-symmetry.  
As a result, the worldvolume coordinates transform under the same supersymmetry as the spacetime coordinates, just as they do (even for the bosonic theories) under the same generalized Lorentz symmetry.

Our new approach differs from previous treatments of $û$-symmetry on branes in that it simultaneously has all the following properties:

\bitem{$\bullet$} It's generated by $ÖPD$ ($+S¯$) \cite{Linch:2016ipx}, as for strings \cite{Siegel:1985xj}, so its parameter is a spacetime spinor.

\bitem{$\bullet$} It's implemented as a first-class constraint, with its own worldvolume Lagrange multiplier/gauge field \cite{Siegel:1985xj}, and is accompanied by the other first-class constraints.

\bitem{$\bullet$} It comes with worldvolume fermionic coordinates $\vartheta_µ$ (``$\rm (supersymmetry)^2Ê$'') \cite{Gates:1985vk, Brooks:1986uh}.

\bitem{$\bullet$} They translate under $û$ symmetry ($¶\vartheta_µ = û_µ$) in the same way the spacetime fermionic coordinates translate under supersymmetry ($¶Î^µ=·^µ$) \cite{Sorokin:1989zi, Sorokin:1988nj}.
\begin{equation*}
\begin{array}{rll}
\hbox{coordinates} & \hbox{bosons} & \hbox{fermions} \\
\hbox{spacetime} & X^m(§,\vartheta) & Î^µ(§,\vartheta) \\
\hbox{worldvolume} & §_{\un\m} & \vartheta_µ
\end{array}
\end{equation*}

There are several advantages over our previous approach:

\bitem{$\circ$} The isotropy group, with generators $S$ and their duals $ê$, is required and determined as part of the current algebra, as a consequence of closure of the super Virasoro algebra.

\bitem{$\circ$} The generalization of the string's first-class constraint system (spacetime supersymmetric Virasoro algebra) \cite{Siegel:1985xj} is manifested.  However, the weight-3 constraint is removed, leaving only constraints of weight 2 (like the usual Virasoro).

\bitem{$\circ$} A fermionic partner to $Î$ is required as a worldvolume gauge field, allowing supersymmetrization of the gauge part of $X$ and thus manifestly covariant worldvolume field strengths.

\bitem{$\circ$} The current $¯$ dual to $D$ is in the dual Lorentz representation (as in S and T-theory); this duality extends to all first-class constraints.

\bitem{$\circ$} The resulting coordinate doubling thus can be removed by supersymmetrizing the selfduality condition.

\bitem{$\circ$} The background gravitino and Lorentz connection are consistent with orthogonality of the vielbein.

\bitem{$\circ$} Earlier linearized results for 3D minimal supergravity \cite{Linch:2015lwa} are confirmed.


\subsection{Notation}

On graded indices, ${}_{[\dots)}$ will stand for graded anti-symmetrization whereas ${}_{(\dots]}$ will denote graded symmetrization. 
(This involves adding terms with coefficients $à$1.)
Indices between bars ${}_{|\dots|}$ are to be left out of (anti-)symmetrization. 
There are the usual implicit relative signs between terms whose indices are in different order (extra $-1$'s from pushing fermionic indices past each other), and Einstein summation has +1's for adjacent indices ordered upper-left and lower-right.

To simplify factors of ``$i$", in our conventions the invariant derivatives are  graded antihermitian, while the currents are graded hermitian. 
So the commutation relations for canonically conjugate variables are now uniformly
\begin{subequations}
\begin{align}
[{\rm p},xÕ & = 1 \\
[\P(1),X(2)Õ & = -i¶(1-2)
\end{align}
\end{subequations}
(i.e., p = $»/»x$, but $\P$ = $-i¶/¶X$) for both bosons and fermions.
(This also implies that the $©$ matrices that appear as structure constants in the supersymmetry algebra are antihermitian, and have an extra ``$-i$" compared to the usual conventions.)
Thus the relation of the action of the fundamental currents $\dd_A$ and invariant derivatives $á_A$ on a function $\phi $ can be expressed as
\begin{align}
\label{E:der}
ñ ­ iÇd§¼ñ^A \dd_AâÜâ[ñ,\phi ] = ñ^A á_A \phi  .
\end{align}
(The Dirac $¶$ functions and integrals $d§$ are over all bosonic and fermionic worldvolume coordinates.)
Currents of the form $\P+»X$ are then normalized without extra powers of 2:  $\P$ to produce the above relation between $\dd$ and $á$, $»X$ so that it appears with the same normalization as $\P$ in a diagonal basis in S-theory.  This also fixes the normalization of the Schwinger term in the current algebra.  (For similar reasons, structure constants $f_{AB}{}^C$ tend to have factors of 2 in them in a real basis, but not in a complex one.  We'll skip this convention in this paper to favor the structure constants, rather than the explicit representation of generators.)

The generators $\S^Œ$ ($\B^Œ$) of $û$-symmetry are generally in the (spacetime) Lorentz (spinor) representation dual to that of the spacetime supersymmetry generators $q_Œ$.  Something analogous (but not so exact) relates generators of spacetime and worldvolume translations \cite{Linch:2015lwa}.  This makes it more natural to label worldvolume derivatives $»^\M=»/»§_\M$ with contravariant indices (superscripts), as opposed to the covariant indices (subscripts) on spacetime derivatives $»_M=»/»x^M$.

We use ``flat" indices $A,B,...$ to label flat superspace, while using ``curved" indices $M,N,...$ to label the coordinate basis (and similarly $\A,\B,...$ and $\M,\N,...$ for the super worldvolume).

(In the appendices we sometimes label $Î$, etc., with underlined spinor indices $\un Œ$, since those ``Lorentz" representations may be reducible.  But in the body of the paper we use simply $Œ$ to improve legibility.)


\subsection{Currents and gaugings}

We will need to define the following related quantities:  (1) worldvolume currents $\dd_A(§)$, defining an affine Lie superalgebra (``weight 1"), (2) spacetime invariant derivatives $á_A$, representing the zero-modes of these currents (ordinary Lie superalgebra), (3) a supersymmetric generalization $\S^\A(§)$ of the Virasoro algebra (including $û$-symmetry), constructed from bilinears of the fundamental currents (``weight 2"), and (4) worldvolume invariant derivatives $\D^\A$, representing the zero-mode part of the generalized Virasoro algebra in terms of worldvolume derivatives.

\begin{equation}
\label{T:notE}
\begin{array}{rll}
& \hbox{current} & \hbox{covariant derivative} \\
\hbox{spacetime} & \dd_A(§) = R_A{}^M(X(§)) \P_M + ... & á_A = R_A{}^M (x) »_M \\
\hbox{worldvolume} & \S^\A = \f14 ú^{BC\A}\dd_B\dd_C & \D^\A = g^\A{}_\M (X(§)) »{}^\M
\end{array}
\end{equation}

There are then three types of currents and the gauge transformations they generate:
$$ \vcenter{
\halign{ #\hfilâ && $#$â \cr
& \hbox{fundamental} & \hbox{Virasoro} & \hbox{Gauss} \cr
generator & \dd_A, á_A & \S^\A, \D^\A & \U_{\vec A} \cr
reparametrizes & X^M, x^M & §_\M &  \cr
parameter & ñ^A(X) & Â_\A(§) & \wedge^{\vec A}(§) \cr
}} $$
The first kind of transformation, generated by the fundamental currents $\dd_A$, is spacetime coordinate transformations, acting on the background fields.  The second kind, generated by the Virasoro operators $\S^\A$, is worldvolume coordinate transformations. 
They are first-class constraints, and so are gauged by Lagrange multipliers that are essentially the worldvolume metric.  Together with selfduality, they imply the existence of the third, ``Gauss" type of gauge transformation, with generators $\U_{\vec A}$.  They are also first-class constraints, and their Lagrange multipliers are the ``$ $-components" of $X$ ($X^M$ being the ``$§$-components").
The bosonic parts of the indices on the currents correspond to particular representations of the exceptional groups E$_{\rm D+1}$:
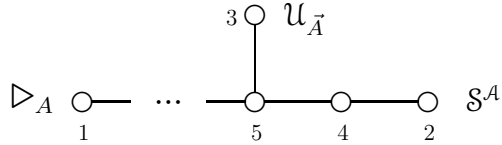
\begin{align*}
\resizebox{2mm}{!}{
$\thicklines
\put(0,0){\circle{10}}
\put(5,0){\line(1,0){35}}
\put(45,0){\circle{10}}
\put(50,0){\line(1,0){35}}
\put(90,0){\circle{10}}
\put(0,5){\line(0,1){35}}
\put(0,45){\circle{10}}
\put(-25,0){\line(1,0){20}}
\put(-50,0){\circle*{2}}\put(-45,0){\circle*{2}}\put(-40,0){\circle*{2}} 
\put(-85,0){\line(1,0){20}}
\put(-90,0){\circle{10}}
\put(-130,-3){\resizebox{9mm}{!}{{$\dd_A$}}}
\put(110,-5){\resizebox{6.5mm}{!}{{$\S^\A$}}}
\put(15,40){\resizebox{7.5mm}{!}{{$\U_{\vec{A}}$}}}
\put(88,-20){$2$}
\put(-15,40){$3$}
\put(43,-20){$4$}
\put(-2,-20){$5$}
\put(-92,-20){$1$}$
} 
\end{align*}

Besides the Virasoro and Gauss constraints $\S$ and $\U$, there are also the ``section" conditions, which result from replacing some of the $\dd$'s in the constraints with $á$'s (zero-modes), acting on functions or their products.  These spontaneously break the symmetry by reducing the dimension of spacetime and (for $\U$) the worldvolume (as the full $\S$ and $\U$ constraints do for the nonzero-modes).


\subsection{Outline}

In the section immediately following we review properties of ordinary Lie superalgebras as they relate to F-theory.
Most of the constraints on the invariant tensors, including those required by the Virasoro algebra, {\it already} appear in the Lie superalgebra.

In the next section we continue our general analysis by discussing properties of the flat superspace current algebras of F-theory: the fundamental (affine Lie) algebra, the Gauss constraint algebra implied by the Virasoro constraints, and the Virasoro algebra itself.  
Then we find explicit expressions for the fundamental currents.  

In the following section we specialize to current algebras of interest, relevant to superstrings.  
We find not only that supersymmetry now {\it requires} the complete set of currents $\dd=SDP¯ê$ considered in previous papers, but  {\it also} an analogous set of super Virasoro generators $\S^\A$ for $\A=\s\d\un\p\varpi\varsigma$, and Gauss constraints $\U=\vec S\vec D\vec P\vec ¯\vec ê$.  We summarize the constraints found earlier, and list various $©$-matrix identities that will be found useful later.  (These identities are proven for various D in the appendices.)

We next find explicit expressions for the constant tensors appearing in these algebras by solving the constraints.
This solution proves the necessity of the set of currents just discussed.
Because of the symmetry relating dual ``Lorentz" representations of currents, a small number of specialized tensors reappear among the components of the general tensors.

We conclude by outlining topics remaining for investigation.


\section{Lie algebra}


\subsection{Representations}

As usual,  the zero-modes of the current algebra define an ordinary Lie superalgebra
\begin{equation}
[G_A,G_BÕ = f_{AB}{}^C G_C
\end{equation}
with (antihermitian) generators $G_A$ and structure constants $f_{AB}{}^C$.
The invariant derivatives for right multiplication on the group element $g(X)$ are
\begin{equation}
\label{E:inv}
á_A = R_A{}^M »_M
\end{equation}
where as usual, for any representation of the group element $g$ and generators $G$
\begin{equation}
\label{E:rielbein}
g^{-1}dg = dX^M R_M{}^A G_A
\end{equation}
The ``flat-space vielbein" $R_A{}^M$ appears in the flat currents in a similar way.
It is in no way related (other than range of indices) to the curved-space vielbein $E_A{}^M$: For example, it does not satisfy an orthogonality relation.  
In particular, the structure constants can be expressed in terms of $R$ in the usual way, without the $úú$ term appearing in the definition of the torsion $T$:
\begin{equation}
\label{E:fdef}
R_{[A|}{}^M (»_M R_{|B)}{}^N) R_N{}^C = f_{AB}{}^C
\end{equation}

Some terminology:  Cosets G/H are conveniently represented by accompanying coordinates for all of G with a gauge group H.  Symmetry is associated with group multiplication on one side, whose corresponding derivatives $ßá$ are then the Killing vectors, while multiplication on the other side is associated with derivatives $á$ that are invariant under the symmetry (commute with the Killing vectors), because of the associativity of group multiplication.  
Gauge invariance of functions of the group is imposed as vanishing of the invariant derivatives $á_{\rm H}$ for the gauge group.  Usually (but not here) a unitary gauge is chosen by fixing the values (``gauging away") the coordinates for H.  If this is done in the coset derivatives $á_{\rm G/H}$, they become no longer invariant under H symmetry transformations, but ``{\it covariant} derivatives".

These symmetry generators are proportional to the invariant derivatives by a factor of the group element $g_M{}^A$ in the adjoint representation, 
\begin{equation}
\label{E:Killing}
ßá_M = g_M{}^A á_A
\end{equation}
(The ``flat" index ``$A$" on $g_M{}^A$ is associated with right multiplication, while the ``curved" index ``$M$'' is associated with left multiplication.)
They can also be expressed directly in terms of partial derivatives with a vielbein for left multiplication as done above for $á$ for right multiplication, using instead $(dg)g^{-1}$.  The above equation then expresses the fact that the ratio of the two vielbeins is given by the group element.

The group space will be identified with (super) spacetime.  This group G is not the one used for the coset of the massless bosonic fields:  It is identified with (first-quantized) coordinates for spacetime, not (second-quantized) fields on spacetime (although they are sometimes interpreted as fields on the worldvolume).  However, the subgroup H is the same, generalized Lorentz symmetry.
(The simplest example of this construction is the one for Minkowski space as Poincar«e/Lorentz.)

We next introduce the worldvolume, by defining new variables $§_\M$ that are a representation of the group $G$.  Thus its partial derivatives $»^\M=»/»§_\M$ are in the dual representation, while invariant derivatives on the worldvolume $\D{}^\A$ can be defined as
\begin{equation}
\label{E:wvd}
\D{}^\A = g^\A{}_\M »^\M
\end{equation}
where $g^\A{}_\M$ is the inverse of the group element $g^\M{}_\A$ in this dual worldvolume representation, and therefore acts as the worldvolume vielbein.
Closure of the algebra of the $û$-symmetry generators will require extension of the worldvolume representation beyond that found in earlier papers. 


\subsection{Invariant tensors}

We also define the ``worldvolume structure constants"
\begin{align}
\label{E:BU}
\ff_{A\B}{}^\C ­ (á_A g^\C{}_\M)g^\M{}_\B 
\end{align}
which can be recognized (see (\ref{E:inv}) and (\ref{E:rielbein})) as the worldvolume representation of the group generators, just as the structure constants $f$ give the adjoint representation:
\begin{equation}
(G_A)_B{}^C = -f_{AB}{}^C¼,â(G_A)^\B{}_\C = -\ff_{A\C}{}^\B
\end{equation}
They thus satisfy the usual Jacobi identities
\begin{subequations}
\begin{align}
\label{E:ff}
f_{[AB}{}^E f_{C)E}{}^D & = 0 \\
\label{E:ff2}
\ff_{[A|\E}{}^\D \ff_{|B)\C}{}^\E + f_{AB}{}^E \ff_{E\C}{}^\D & = 0
\end{align}
\end{subequations}
which can also be obtained from the curl of (\ref{E:fdef}) and (\ref{E:BU}).
The group element $g$ in the two representations can be expressed, e.g., by the usual exponential parametrization $g=e^{X^A G_A}$.

The invariant tensor $ú^{AB\C}$ will be used to construct the Virasoro operators, while $ú_{AB\C}$ will appear in the current algebra.  These tensors act as Clebsch-Gordan-Wigner coefficients relating the spacetime ($A$) and worldvolume ($\A$) representations in the  (graded) symmetric part of the product $(A°B)_S=\C + ...$ and its dual.  Their invariance under infinitesimal group transformations yields the identities
\begin{subequations}
\begin{align}
\label{E:down}
f_{A(B}{}^E ú_{C]E\D} & = \ff_{A\D}{}^\E ú_{BC\E}\\
\label{E:up}
f_{AE}{}^{(B}ú^{C]E\D} & = \ff_{A\E}{}^\D ú^{BC\E}
\end{align}
\end{subequations}
These are the generalizations of the $f$ symmetry conditions of T-theory, broken by $\ff$.

From separating the former into its totally symmetric and mixed symmetry pieces we have
\begin{subequations}
\begin{align}
\label{E:downer}
ú_{(AB|\E}\ff_{|C]\D}{}^\E & = 0 \\
\label{E:downer2}
ú_{E(A|\D}f_{|B]C}{}^{E} & = \f13\eta_{C (A|\E} \ff_{|B]\D}{}^\E -  \f23\eta_{AB\E} \ff_{C\D}{}^\E
\end{align}
\end{subequations}
Since it has mixed symmetry, the last can also be written as
\begin{equation}
\label{E:downer2b}
ú_{E[A|\D}f_{|B)C}{}^E - 2ú_{CE\D}f_{AB}{}^E = ú_{C[A|\E}\ff_{|B)\D}{}^\E
\end{equation}
From tracing (\ref{E:up}) we have
\begin{equation}
\label{E:uptr}
f_{AB}{}^B = 0âÜâ\ff_{B\C}{}^\A ú^{AB\C} = 0
\end{equation}

Defining ``$ $" as an arbitrary (bosonic) direction in the worldvolume space, we can use $ú^{AB }$ as an (almost) ordinary metric:
We can then define a type of ``duality" symmetry (as in T-theory) as the ``reflection" symmetry 
\begin{equation}
G_A ª ú^{AB }G_B
\end{equation}
relating pieces of $G_A$ in dual ``Lorentz" representations.

We will find a similar duality symmetry on the worldvolume derivatives,
\begin{equation}
\D^\A ª ú_{\A\B}\D^\B
\end{equation}
for a 
metric 
\begin{equation}
ú_{\A\B} = ú^{CD }ú^{EF }ú_{CE\A}ú_{DF\B}
\end{equation}
(The ``bosonic" part is the ``Minkowski" metric.
Similar remarks apply to the CGW coefficients of the Gauss constraints.)
The combined dualities relate (\ref{E:down}) to (\ref{E:up}).


\section{Flat currents}


\subsection{Fundamental}

We now give a semi-explicit representation for the currents of flat superspace.
These are to be used as a basis for the currents of curved superspace.  The flat currents are a generalization of those appearing in 2D nonlinear $§$-models defined on compact group spaces \cite{Witten:1983ar} and their generalization to the noncompact groups of superstring theory \cite{Green:1989nn, Siegel:1994xr, Sakaguchi:1998kk}.

The fundamental currents $\dd_A$ satisfy the bracket rule
\begin{align}
\label{E:CurrentAlgebraFlat}
i[\dd_A (1), \dd_B (2) \} & =  ¶(1-2) f_{AB}{}^C \dd_C\ + \eta_{AB\C} \D{}^\C_{2-1} ¶(1-2)
\end{align}
where $\O_{2-1}­\O(2)-\O(1)$.  The currents can be expressed in the form (introducing the spacetime 2-form $B$)
\begin{equation}
\label{E:nacho}
\dd_A = R_A{}^M\P_M + (B_{AB\C} + ú_{AB\C})(\D{}^\C X^M)R_M{}^B
\end{equation}
Just as for $R$ vs.¼$E$, the ``flat-space differential form" $B$ is not related to the curved spaced one, which appears only as a part of $E$ upon solving its orthogonality constraint.  

In addition to the selfdual currents $\dd_A$ we have the antiselfdual ones ${\widetilde\dd}_A$, which are normally ignored as second-class constraints,
\begin{equation}
\label{E:SD}
{\widetilde\dd}_A = R_A{}^M\P_M + (B_{AB\C} - ú_{AB\C})(\D{}^\C X^M)R_M{}^B = 0
\end{equation}
(In relating to the Lagrangian formalism, $\P+B\D X¾ÀX$; the selfdual and antiselfdual currents are $¾ÀXà\D X$.)
The symmetry currents are proportional to the anti-selfdual currents by a factor of the group element $g_M{}^A$, 
\begin{equation}
\label{E:sym}
{ß\dd}_M = g_M{}^A{\widetilde\dd}_A
\end{equation}
so the selfduality constraints can be identified as vanishing of some symmetry currents.  
The Killing vectors (\ref{E:Killing}) then follow from the zero-modes,
\begin{equation*}
ßá_M = g_M{}^A á_A
\end{equation*}
since the zero-modes of $\dd$ and $\widetilde\dd$ are the same.
Furthermore, in an appropriate gauge $g_M{}^A$ can be taken as triangular with respect to engineering dimension, so imposing the upper half of one as first class can be interpreted as imposing the upper half of the other.
The $\S$-sectioning constraints, quadratic in these zero-modes, can then be interpreted as covariant first-class versions of the zero-mode part of the selfduality condition, effectively imposing half the Killing vectors as constraints.


\subsection{Gauss}

The fundamental currents are related to spacetime invariant derivatives directly as (\ref{E:der})
\begin{equation}
i[ \dd_A (1) , \phi (X(2)) Õ = ¶(1-2) á_A \phi 
\end{equation}
However, the Virasoro currents (\ref{T:notE})
\begin{equation}
\S^\A = \f14 ú^{BC\A}\dd_B\dd_C
\end{equation}
are related more subtly to the worldvolume invariant derivatives:
We require
\begin{equation}
i[ \S^\A, \phi (X) Õ = ¶ \D^\A \phi 
\end{equation}
Since explicit evaluation gives
\begin{equation}
i[ \S^\A, \phi (X) Õ = ¶ üú^{BC\A}\dd_B á_C \phi 
\end{equation}
and
\begin{equation}
\D^\A \phi (X(§)) = (\D^\A X^M)»_M \phi 
\end{equation}
while selfduality (\ref{E:SD}) implies
\begin{equation}
\widetilde\dd_A = 0âÜâ\dd_A = 2ú_{AB\C}(\D^\C X)^B¼,â(\D^\A X)^A ­ (\D^\A X^M)R_M{}^A
\end{equation}
this requires ``cancelation" of upper and lower $ú$'s by the constraint that generates the Gauss gauge transformation:
\begin{equation}
\label{E:U}
\U_A^\A ­ U^{B\A}_{A\B}\dd_B\D^\B = 0¼,â
	U^{B\A}_{A\B} ­ ¶^B_A ¶^\A_\B - ú^{CB\A} ú_{CA\B}
\end{equation}

Here the tensor $U$ takes a simple form in terms of $ú$, but requires some CGW coefficients to factorize on the irreducible ``$\vec AÊ$" representation:
\begin{equation}
\label{E:CGW}
U^{B\A}_{A\B} = c^{\A\vec C}_A c^B_{\B\vec C}â
Üâ\U_A^\A = c^{\A\vec C}_A \U_{\vec C}¼,â\U_{\vec C} = c^B_{\B\vec C}\dd_B\D^\B
\end{equation}

This constraint is a result of the spacetime coordinates being gauge fields on the worldvolume: e.g., $X(§, )$ resembles a worldvolume differential form.  In the Hamiltonian formalism in which our present discussion takes place, separation of $ $ and $§$ (and the corresponding reduction in manifest symmetry) has separated these gauge fields $X^{\un M}$ into their ``$ $-components" $X^{\vec M}$, which act as Lagrange multipliers for the corresponding ``Gauss constraints" $\U$, and their purely ``$§$-components" $X^M$, which survive as propagating on the worldvolume in the temporal gauge \cite{Linch:2015fya}.

Thus
\begin{equation}
\D^\A \phi (X(§)) = üú^{BC\A}\dd_B á_C \phi 
\end{equation}
and in particular,
\begin{equation}
\label{E:SD2}
(\D^\A X)^A = üú^{AB\A}\dd_B
\end{equation}
(again imposing selfduality and the Gauss constraint).  This is the only form in which selfduality (combined with Gauss constraints) will need to be imposed in the following manipulations.


\subsection{Virasoro}

The Virasoro algebra transforms the currents and vice versa as, applying (\ref{E:up}),
\begin{equation}
\label{E:Vf}
i[ \S^\A(1),\dd_A(2)Õ = 
	-¶\ff_{A\B}{}^\A\S^\B + ü\dd_A(1)\D_{2-1}^\A ¶ -ü(¶\U^\A_A + \U^\A_{A,2-1}¶)
\end{equation}
where $¶\U­¶U(\D\dd)$, $\U ¶ ­ U\dd(\D ¶)$%
.  The Virasoro algebra is then (classically)
\begin{align}
i[ \S^\A (1), \S^\B (2) Õ = ¶üú^{CD[\A}\ff_{C\E}{}^{\B)}\dd_D\S^\E 
+ [ \S^\A ü(1+2) \D^\B(2) ¶ +\f14 ú^{AB\B}\dd_A(2)\U^\A_B(1) ¶ - {\scriptstyle{\Aª\B\choose 1£2}} ]
\end{align}
where $\Sü(1+2) ­ ü[\S(1)+\S(2)]$.
(Previously we considered also ``$\V$" and ``$\W$" constraints, which we can assume as derived from $\U$ and $\widetilde\dd$.)
This should be compared with, applying selfduality as (\ref{E:SD2}), but not (\ref{E:up}),
\begin{equation}
[ \D^\A , \D^\B Õ = üú^{CD[\A}\ff_{C\E}{}^{\B)}\dd_D\D^\E
\end{equation}

Note that this algebra resembles the algebra of first-class constraints for S-theory \cite{Siegel:1985xj}, if we introduce $f$, $\ff$, and $ú^{AB\C}$, but only $ú_{AB}$ (not $ú_{AB\C}$).  Our previous (super) F-theory paper \cite{Linch:2016ipx} corresponded to using a bosonic worldvolume index $\c$ in $ú_{AB\c}$ and thus $\D^\a$, but an extended ``super" worldvolume index $\A$ in $\ff_{A\B}{}^\C$ and $ú^{AB\C}$, and thus in the complete algebra of Virasoro-related first-class constraints $\S^\A$, while $\ff_{A\b}{}^\c=0$.  In principle, even after extending the worldvolume to a superspace, there need not be an identification between the $ú$'s and $\ff$'s that appear in the $\D$ and $\S$ algebras.  But we'll see later that this identification is required by the explicit form of the commutation relations for the algebras relevant to F-theory.  At this point in our general discussion such an identification already seems natural.


\subsection{2-form}

Explicit evalution of the $\dd$ algebra (\ref{E:CurrentAlgebraFlat}) by substituting (\ref{E:nacho}) yields
\begin{equation}
\label{E:delta}
i[\dd_A (1), \dd_B (2) \} =  ¶ f_{AB}{}^C \dd_C\ + \eta_{AB\C} \D{}^\C_{2-1} ¶ + ¶ ë_{ABC\D}(\D^\D X)^C
\end{equation}
where the non-closure term $¶ë\D X$ is given by, separating $ë=A+M$ the totally antisymmetric ($A$) and mixed symmetry ($M$) parts in $ABC$,
\begin{align}
A_{ABC\D} & = H_{ABC\D} + \f16 f_{[AB|}{}^E ú_{E|C)\D}\\
M_{ABC\D} & = \f23(ú_{E[A|\D}f_{|B)C}{}^E - 2ú_{CE\D}f_{AB}{}^E - ú_{C[A|\E}\ff_{|B)\D}{}^\E) 
	- \f13 ú_{C[A|\E}\ff_{|B)\D}{}^\E
\end{align}
Here the field strength $H$ of $B$ is defined in the usual way:  
With curved indices it's simply the (graded) curl of $B$,
\begin{equation}
H_{MNP\Q} = ü»_{[M}B_{NP)\Q}
\end{equation}
and thus flattening the indices
\begin{equation}
H_{ABC\D} = üá_{[A}B_{BC)\D} -üf_{[AB|}{}^E B_{E|C)\D} +ü\ff_{[A|\D}{}^\E B_{|BC)\E}
\end{equation}

We have already assumed constant $ú$,
\begin{equation}
á_A ú_{BC\D} = 0
\end{equation}
The first part of $M$ vanishes by the algebraic constraint (\ref{E:downer2b})%
.  We can then obtain closure of the $\dd$ algebra by applying the selfduality constraint in the form (\ref{E:SD2}) to produce an $ú$ from $\D X$. 
The $A$ contribution and that from the rest of $M$ can be made to vanish separately:
For the latter we need the final constraint
\begin{equation}
\label{E:wot}
ú^{FC\D} ú_{C[A|\E}\ff_{|B)\D}{}^\E = 0
\end{equation}
For the former we want to solve
\begin{equation}
\label{E:Hsol}
ú^{FC\D} (H_{ABC\D} + \f16 f_{[AB|}{}^E ú_{E|C)\D}) = 0
\end{equation}
With the help of (\ref{E:downer2b}) and (\ref{E:wot}), this can also be written in a form similar to S-theory as
\begin{equation}
H_{ABD\E} ú^{CD\E} = - k f_{AB}{}^C
\end{equation}
where $k$ is the normalization appearing in
\begin{equation}
\label{E:norm}
ú_{AC\D} ú^{BC\D} = k ¶_A^B
\end{equation}

A solution for (\ref{E:Hsol}) can be found in a convenient gauge where $B$ is chosen constant:
Then dropping the $áB$ term expresses $H$ in terms of just the constants $B,f$, and $\ff$.
The result is \cite{Hatsuda:2015cia}
\begin{equation}
\label{E:B}
B_{AB\C} = {w_A - w_B\over w_A + w_B}Êú_{AB\C}â(0¼\hbox{if}¼w_A = w_B = 0)
\end{equation}
(indices on $w$ not summed) where $w_A$ is the spacetime scaling weight (engineering dimension) associated with the current $\dd_A$.  Similarly, we can associate a weight $w_\A$ to $\S^\A$ and $\D^\A$; then the dimensionlessness of the constant tensors implies
\begin{align}
ú_{AB\C} : &âw_A + w_B - w_\C = 0 \\
f_{AB}{}^C : &âw_A + w_B - w_C = 0 \\
\ff_{A\B}{}^\C : &âw_A - w_\B + w_\C = 0 \\
c^{\B\vec C}_A : &âw_A + w_\B - w_{\vec C} = 0
\end{align}
where
\begin{equation}
w_A ³ 0¼,âw_\A ³ 1¼,âw_{\vec A} ³ 2
\end{equation}
(for reasons explained in the next section).

The constraint is then found to be satisfied using again (\ref{E:downer2b}) and (\ref{E:wot}), as well as the symmetric analog of (\ref{E:wot}):
\begin{equation}
\label{E:wot2}
ú^{FC\D} ú_{C(A|\E}\ff_{|B]\D}{}^\E = 0
\end{equation}
However, this already vanishes, using the Lie algebra identities.  The explicit (gauge-inde\-pend\-ent) form for $H$ found from this constant-gauge $B$ is
\begin{equation}
\label{E:H}
H_{ABC\D} + \f16 f_{[AB|}{}^E ú_{E|C)\D} = 
	(w_B - w_C)\left( {1\over w_B + w_C} - \f23 {1\over w_A + w_B + w_C} \right) \ff_{A\D}{}^\E ú_{BC\E} 
	+ \hbox{cyc.¼perm.}
\end{equation}

The symmetry currents (\ref{E:sym}) should also commute with the covariant currents:
\begin{equation}
[\dd_A,ß\dd_MÕ = 0
\end{equation}
Explicit evaluation shows this to be the case with the constraints above.  

As a check of the above solution of the algebra, we can also solve its Jacobi identities.
We are allowed to apply selfduality (\ref{E:SD2}) to evaluate these identities, but not on the $[\dd,\ddÕ$ algebra itself beforehand, which we therefore write as (\ref{E:delta}) (but we can drop the part of $ë$ proportional to (\ref{E:downer2b})).
The Jacobi identities then imply (\ref{E:ff}) and (\ref{E:downer2b}).


\section{Specialization}


\subsection{Supersymmetry}

The supersymmetrization of the F-theory current algebra presented in some of our previous papers \cite{Linch:2015fca,Linch:2016ipx}
has an inconsistency:  It missed the requirement that the worldvolume also be generalized to a superspace.  This result could have been guessed from the close relation between it and spacetime, or from the fact that the Virasoro algebra in S-theory is generalized to a type of superalgebra by the completion of first-class constraints. 

This inconsistency would appear when solving the orthogonality conditions for a massless background.  Although we do not consider such backgrounds explicitly in this paper, some of the conditions on the torsion are essentially just first derivatives of (integrability conditions for) the orthogonality conditions.  In particular, we examine conditions (including Bianchi identities) on the flat version of the torsions, the structure constants $f$ and $\ff$.

Unlike the previous sections, we now specialize the current algebra to supersymmetric algebras of interest.
We start from the minimal version of the previous papers:  the currents $D,P,¯$ and bosonic worldvolume covariant derivative $\p$.  As we solve the constraints, we will find not only the necessity of new worldvolume (super)coordinates, but also a change (simplification) in indices for $¯$.  (The latter was hinted by our original treatment of F-theory superspace \cite{Linch:2015lwa}.)

Ultimately, we find the complete set of currents $\dd=SDP¯ê$ (as in S and T-theory), worldvolume invariant derivatives $\D=\s\d\un\p\varpi\varsigma$, and Gauss constraints $\U=\vec S\vec D\vec P\vec ¯\vec ê$.
The currents $SDP¯ê$ have weights $w_A=0,ü,1,\f32,2$.
Furthermore, $\p$ is found to extend to $\un\p$ (details below).
There is a correspondence between $\s\d\un\p\varpi\varsigma$ and $SDP¯ê$, as indicated by the lightcone formalism, since 
\begin{equation}
\S^\A ¾ ú^{Ab\A}P_b \dd_A ® ú^{A-\A}P_-\dd_A
\end{equation}
relates $\S^\A$ and $\dd_A$ with engineering dimensions $w_\A=w_A+1$.
Similar remarks apply between $\vec S\vec D\vec P\vec ¯\vec ê$ and $\s\d\un\p\varpi\varsigma$, relating $\U_{\vec A}$ and $\D^\A$ with $w_{\vec A}=w_\A+1$:
\begin{equation}
\U_{\vec A} ® c^-_{\B\vec A}P_-\D^\B
\end{equation}
We can also relate $\dd$ and $\U$ directly by picking out the component $\p^-$ of $\un\p$ that survives reduction to S and T-theory:
\begin{equation}
\U_{\vec A} ® c^A_{-\vec A}\p^-\dd_A
\end{equation}
with $w_{\vec A}=w_A+2$.

There is then the pairing of coordinates $X^{\un M}=(X^M,X^{\vec M})$, of dynamic coordinates $X^M$ with Lagrange multipliers $X^{\vec M}$ coming from $\dd$ and $\U$ for each engineering dimension (as found in previous papers for $P$ \cite{Linch:2015fya, Linch:2015qva,Linch:2015fca,Linch:2016ipx}), as representations of the ``Lorentz" symmetry of the full worldvolume (including $ $).


\subsection{Constraints}

We have found various types of identities required of the constant tensors, which we collect here.
From Lie algebra we have from the commutation relations 
\begin{subequations}
\begin{align}
\label{E:fff}
f_{[AB}{}^E f_{C)E}{}^D & = 0 \\
\label{E:fff2}
\ff_{[A|\E}{}^\D \ff_{|B)\C}{}^\E + f_{AB}{}^E \ff_{E\C}{}^\D & = 0
\end{align}
\end{subequations}
and from the invariance of the CGW coefficients
\begin{subequations}
\begin{align}
\label{E:stronger}
f_{AE}{}^{(B}ú^{C]E\D} & = \ff_{A\E}{}^\D ú^{BC\E} \\
\label{E:ffeta}
f_{A(B}{}^E ú_{C]E\D} & = \ff_{A\D}{}^\E ú_{BC\E}
\end{align}
\end{subequations}

From solving the fundamental current algebra:
\begin{align}
ú^{FC\D} ú_{C[A|\E}\ff_{|B)\D}{}^\E & = 0 
\end{align}

These constraints also imply some $úúú$ identities (discussed for the bosonic case in \cite{Linch:2016ipx}).
They will be analyzed in a future paper.



\subsection{\texorpdfstring{\boldmath$©$}{\textgamma} matrix identities}

We will also use some generic Dirac-matrix identities: 
There is the relation between spacetime matrices $©^a_{Œº}$, $©^{aŒº}$ and worldvolume matrices $ý_{\un\a º}{}^©$, and its dual
\begin{subequations}
\label{E:gama}
\begin{align}
\label{E:gamma}
©^{(a}_{Œ©}©^{b)©º} & = - ú^{ab\un\c}ý_{\un\c Œ}{}^ºâ(ý_{ Œ}{}^º ¾ ¶_Œ^º)\\
\label{E:gammma}
©_{(a|Œ©}©_{|b)}^{©º} & = - ú_{ab\un\c}ý_Œ^{\un\c º}â(ý_Œ^{ º} ¾ ¶_Œ^º)
\end{align}
\end{subequations}
The $©$-matrices are given by the supersymmetry algebra, essentially by knowing the range of the indices $Œ$ and $a$.  The above identities then fix the range of index $\un\a$, as well as determining the $ý$-matrices and $ú^{PP\un\p}$ (and its dual).  
(They appear in (\ref{E:Gamma}) and (\ref{E:mama}).  Cases are examined in Appendix \ref{A:gama}.)
These require the extension $\p£\un\p=(\p, )$.  A closely related identity (appearing in (\ref{E:ddual}) and Appendix \ref{A:more}) is
\begin{equation}
©^a_{ú(·} ý^\b_{½)}{}^ú = ú^{ah\b}©_{h·½}
\end{equation}

Another important identity (appearing in (\ref{E:Lorentz})) relates the $©$-matrices to the group generators $S$ and their representations:
\begin{equation}
©_a^{Œ©}©^b_{Œº} - ©_{aŒº}©^{bŒ©} = f_{Sa}{}^b ©^{S©}_º
\end{equation}
Solving this identity (in Appendix \ref{A:Lorentz}) not only determines the vector ($f_{Sa}{}^b$) and spinor ($©^{S©}_º$) representations, but also the group itself.

We also have the Fierz identities (see (\ref{E:fierz}))
\begin{equation}
\label{E:Fierz}
©^a_{(Œº}©^b_{©)¶}ú_{ab\un\c} = 0
\end{equation}
which can be satisfied in D = 3,4,6,10.  
However, we sometimes need to supplement $X$ with the worldvolume scalars $Y$ (and their worldvolume duals $\widetilde Y$) to raise D to those critical values.  (See Appendix \ref{A:Fierz} for more details.)


\section{Constraint solution}
\label{S:BI}


\subsection{Outline}


In this section we solve many of the constraints collected in the previous section (with proofs for various D left for the appendices).  But first we give a listing of components of these constraints, and the results found from each one:

\begin{equation*}
\begin{array}{lllll}
\hbox{constraint} & \hbox{gives} & \hbox{equation} & \hbox{``dual"} & \hbox{appendix} \\
\hline
(ff)_{DDD}^¯ & \hbox{Fierz}, ¯ & (\ref{E:fierz}) & & \hbox{\ref{A:Fierz}} \\
(fú)_D^{P¯\p} & \varpi & (\ref{E:dual}) & (\ref{E:duel}) & \hbox{\ref{A:more}} \\
(fú)_P^{¯¯\p} & \varsigma & (\ref{E:vars}) & (\ref{E:duel2}) & \hbox{\ref{A:more}} \\
(fú)_P^{P¯\d} & S & (\ref{E:S}) & (\ref{E:duwop}) & \hbox{\ref{A:Lorentz}} \\
(fú)_{PPD\varpi} & ê & (\ref{E:duwop}) & (\ref{E:S}) & \hbox{\ref{A:Lorentz}} \\
(fú)_D^{PP\d} & \un\p,ý_{\un\p},ú^{PP\un\p} & (\ref{E:Gamma}) & (\ref{E:mama}) & \hbox{\ref{A:gama}} \\
(fú)_{DDP\p} & ¯,\d & (\ref{E:duel}) & (\ref{E:dual}) & \hbox{\ref{A:more}} \\
(fú)_{PDD\p} & \s & (\ref{E:duel2}) & (\ref{E:vars}) & \hbox{\ref{A:more}} \\
(fú)_{DDD\d} & \s & (\ref{E:smore}) & (\ref{E:doowop}) & \hbox{\ref{A:Fierz}} \\
(fú)_D^{¯¯\varpi} & \varsigma & (\ref{E:doowop}) & (\ref{E:smore}) & \hbox{\ref{A:Fierz}} \\
(fú)_{DPP\varpi} & \un\p,ý^{\un\p},ú_{PP\un\p} & (\ref{E:mama}) & (\ref{E:Gamma}) & \hbox{\ref{A:gama}}
\end{array}
\end{equation*}
where $(ff)_{ABC}^E$ refers to (\ref{E:fff}), $(fú)_A^{BC\E}$ to (\ref{E:stronger}), and $(fú)_{ABC\E}$ to (\ref{E:ffeta}).

The basic idea is that, starting from just the $ÓD,DÕ¾P$ part of the Lie algebra, the structure of the theory is tight enough to determine the rest of the Lie algebra ($S¯ê$), the worldvolume ($\s\d\un\p\varpi\varsigma$), and the various current algebras (fundamental, Virasoro, Gauss).  In particular, this includes finding the Lorentz group H$_{\rm D}$ ($S$) and its representations by all these quantities (at least for D $²$ 5).


\subsection{Cases of \texorpdfstring{\boldmath$¯$}{\textOmega}}

We first look at the only identity (\ref{E:fff}) involving only $f$ and neither $\ff$ nor $ú$ (and thus the only Bianchi identity not affected by the change in worldvolume).
Since the only nonvanishing $f$'s involving only $DP¯$ are
\begin{equation}
f_{DD}{}^P = ©^c_{Œº} ,âf_{DP}{}^¯ 
\end{equation}
the only nontrivial such identity is $(ff)_{DDD}{}^¯$:
\begin{equation}
\label{E:fierz}
f_{(DD}{}^P f_{D)P}{}^¯ = 0
\end{equation}
(Sometimes we will use the symbols for the currents/worldvolume invariant derivatives in place of their indices.
Actually, the necessity of $f_{DP}{}^¯$, and of $¯$ itself, follows from the Bianchi identity (\ref{E:OMGa}) below.)
Its solutions follow from the Fierz identities (\ref{E:Fierz}), 
which suggest two possibilities, using either the $§$ or $ $ parts of the identities.  
The former was considered previously \cite{Linch:2015fca}:  It leads to the solution
\begin{equation}
¯ = ¯^{Œ\a}¼,âf_{DP}{}^¯ = ©^d_{Œ©}ú_{bd\c}
\end{equation}
The latter gives instead
\begin{equation}
¯ = ¯^Œ¼,âf_{DP}{}^¯ = ©^d_{Œ©}ú_{bd }
\end{equation}
This $¯$ has the same indices as in S-theory and T-theory.  As a result, its gauge parameter $ñ^¯=ñ_Œ$ agrees with that found from the analysis of the superspace of (linearized) 3D F-theory \cite{Linch:2015lwa}.

Then the only nonvanishing $f$'s and $ú$'s at this point are
\begin{equation}
f_{DD}{}^P = ©^c_{Œº}¼,âf_{DP}{}^¯ = ©^d_{Œ©}ú_{bd\c}¼\hbox{(for }¯^{Œ\a}\hbox{)âor}â©^d_{Œ©}ú_{bd }¼\hbox{(for }¯^Œ)
\end{equation}
\begin{equation}
ú_{PP\p} = ú_{ab\c}¼,âú_{D¯\p} = ¶_Œ^º ¶_\c^\b¼\hbox{(for }¯^{Œ\a}\hbox{)âor}âý_{\c Œ}{}^º¼\hbox{(for }¯^Œ)
\end{equation}
(as follows from dimensional analysis and ``Lorentz" invariance).


\subsection{Virasoro}

We now consider the consequences of the identity needed for the Virasoro algebra (\ref{E:stronger}).
Because this uses $ú^{AB\C}$ rather than $ú_{AB\C}$, it will more easily constrain currents of higher engineering dimension rather than lower.
But they also resolve the $¯$ ambiguity of the previous subsection.

We examine the $(fú)_D^{P¯\p}$ identity, needed for $[\S^\p,D]$.
If we still consider just $DP¯$ fundamental currents, and $\S^\p$ Virasoro, ignoring $\ff$, neither type of $¯$ offers a solution.
So we include the $\ff$ term, which in this case requires the existence of the $\varpi$ Virasoro $P¯$:
\begin{equation}
\label{E:dual}
- f_{DD}{}^P ú^{¯ D\p} + f_{DP}{}^¯ ú^{PP\p} = -\ff_{D\varpi}{}^\p ú ^{P¯\varpi}
\end{equation}
(paying careful attention to implicit signs from relative ordering of fermionic indices).
Then for the two cases of $¯$ we have (with $\varpi_\a^Œ$ for $¯^{\a Œ}$, and $\varpi^Œ$ for $¯^Œ$)
\begin{subequations}
\begin{align}
- ©^a_{Œº}(¶^\b_\d ¶^º_¶) + (ú_{be\d}©^e_{Œ¶})ú^{ab\b} & = - (¶^\b_\c ¶^©_Œ)(¶^\c_\d ©^a_{©¶}) \\
\label{E:ddual}
- ©^a_{Œº} ý^\b_¶{}^º + ©_{bŒ¶}ú^{ab\b} & = ý^\b_Œ{}^º ©^a_{¶º}
\end{align}
\end{subequations}
where we have used (again from dimensional analysis and Lorentz invariance, up to coefficients)
\begin{subequations}
\begin{align}
\ff_{D\varpi}{}^\p & = ¶_\b^\c ¶_Œ^º¼,âú^{P¯\varpi} = ¶^\c_\b ©^a_{º©} \\
\ff_{D\varpi}{}^\p & = - ý_Œ^{\c º}¼,âú^{P¯\varpi} = ©^a_{º©}
\end{align}
\end{subequations}
The former ($¯^{\a Œ}$) case is inconsistent, while the latter ($¯^Œ$) case works.

Similarly, we find the necessity of the $\varsigma$ Virasoro $¯¯$ from $(fú)_P^{¯¯\p}$:
\begin{equation}
\label{E:vars}
f_{DP}{}^{[¯}ú^{¯]D\p} = \ff_{P\varsigma}{}^\p ú^{¯¯\varsigma}
\end{equation}
or more explicitly
\begin{equation}
\label{E:moreexplicitly}
©_{aŒ[º}ý^{\b Œ}_{©]} = \ff_{a\varsigma}{}^\b ú_{º©}^\varsigma
\end{equation}
which again could not be satisfied without the right-hand side.

The existence of $\varpi$ Virasoro and of $\d$ Virasoro = $û$ symmetry (see below) also requires $S$ in the fundamental algebra:  From the $(fú)_P^{P¯\d}$ identity,
\begin{equation}
\label{E:S}
-f_{PD}{}^¯ ú^{PD\d} + f_{PS}{}^P ú^{¯S\d} = \ff_{P\varpi}{}^\d ú^{P¯\varpi}
\end{equation}
we find
\begin{equation}
\label{E:Lorentz}
©_{aŒº}©^{bŒ©} + f_{Sa}{}^b ©^{S©}_º = ©_a^{Œ©}©^b_{Œº}
\end{equation}
\begin{equation}
ú^{PD\d} = ©^{aº©}¼,â\ff_{P\varpi}{}^\d = ©_a^{º©}
\end{equation}
which identifies $f_{Sa}{}^b$ and $©^{S©}_º$ as the usual vector and spinor representations of $S$, without which this identity could not be satisfied.  (Any one term requires all three.  $S$ then requires $ê$ in the algebra in the same way that $D$ required $¯$.)
This identity is almost identical to the ``dual" identity $(fú)_{PPD\varpi}$,
\begin{equation}
\label{E:duwop}
 f_{PD}{}^¯ ú_{P¯\varpi} + f_{PP}{}^ê ú_{Dê\varpi} = -\ff_{P\varpi}{}^\d ú_{PD\d}
\end{equation}
up to raising/lowering an $a$ index with $ú_{ab }$, under the identifications
\begin{equation}
ú_{P¯\varpi} £ ú^{PD\d}¼,âf_{PP}{}^ê £ -f_{PS}{}^P¼,âú_{Dê\varpi} £ ú^{¯S\d}¼,âú_{PD\d} £ ú^{P¯\varpi}
\end{equation}

The case $(fú)_D^{PP\d}$ yields the $©$-matrix relation (\ref{E:gamma}), as
\begin{equation}
\label{E:Gamma}
f_{DD}{}^{(P}ú^{P)D\d} = \ff_{D\un\p}{}^\d ú^{PP\un\p}
\end{equation}
Note that this requires $\ff_{D }{}^\d±0$.


\subsection{\texorpdfstring{\boldmath$û$}{\textkappa}-symmetry}

Whereas (\ref{E:stronger}) implied the existence of the higher-dimensional $\varpi$ ($P¯$) and $\varsigma$ ($¯¯$), (\ref{E:ffeta}) requires the lower-dimensional $\d$ ($PD$) of $û$-symmetry, and $\s$ ($DD$).  
Of these identities, the only nontrivial ones for just $DP¯\p$ are $(fú)_{DDP\p}$ and $(fú)_{PDD\p}$:  If we ignore the $\ff$'s,
\begin{subequations}
\begin{align}
\label{E:OMGa}
f_{DD}{}^P ú_{PP\p} - f_{DP}{}^¯ ú_{D¯\p} = 0 \\
\label{E:OMGb}
f_{P[D}{}^¯ ú_{D]¯\p} = 0
\end{align}
\end{subequations}
where the latter is simply the antisymmetrization of the former.  In fact, the former is the one that shows the necessity of the existence of $¯$ (in T-theory and S-theory as well).  
For $¯^{Œ\a}$ these identities would be satisfied.  

But we saw in the previous subsection the requirement of $¯^Œ$ instead.
There the worldvolume needs again to be extended:  For $¯^{Œ\a}$ we could neglect $\ff$ terms because of the vanishing of the $ú$'s multiplying them.  However, the former identity, more explicitly
\begin{equation}
©^b_{Œº}ú_{ab\a} - ©_{aŒ©}ý_{\a º}{}^© = 0
\end{equation}
has no solution even in the 3D case, where in normal Dirac $©$-matrix notation it is something of the form 
\begin{equation}
ý^3 - ý^2 ý = 0
\end{equation}
(Similar remarks apply to the latter identity.)

If we want to keep the spacetime unmodified, the simplest modification is to introduce an $ú$ carrying $PD$ indices:  This corresponds to adding the $û$-symmetry generator $\S^\d=ÖPD$ (and its associated worldvolume covariant derivative $\d$) to the Virasoro generators $\S^\p=üP^2+¯ D$.  This Bianchi identity then becomes
\begin{equation}
\label{E:duel}
f_{DD}{}^P ú_{PP\p} - f_{DP}{}^¯ ú_{D¯\p} = - \ff_{D\p}{}^\d ú_{DP\d}
\end{equation}
(Note that this equation is in some sense ``dual" to (\ref{E:dual}), exchanging currents with dual currents when raising/lowering indices.)
More explicitly
\begin{equation}
\label{E:dddual}
©^e_{Œº}ú_{ce\ff} - ©_{cŒ·}ý_{\ff º}{}^· = ý_{\ff Œ}{}^© ©_{cº©}
\end{equation}
(dual to (\ref{E:ddual})), where we have set
\begin{equation}
ú_{DP\d} = ©_{bŒ©}¼,â\ff_{D\p}{}^\d = - ý_{\b Œ}{}^©
\end{equation}
as the only available invariant tensors, up to normalization.  
This equation is satisfied.

However, this $\ff$ does not appear in the latter identity (\ref{E:OMGb}), which has instead on its right-hand side:  By the dual to (\ref{E:vars}),
\begin{equation}
\label{E:duel2}
f_{P[D}{}^¯ ú_{D]¯\p} = ú_{DD\s}\ff_{P\p}{}^\s
\end{equation}
or more explicitly
\begin{equation}
\label{E:whatever}
©_{c·[Œ|}ý_{\ff |º]}{}^· = ú_{Œº\s}\ff_{c\ff}{}^\s
\end{equation}
introducing a new Virasoro constraint $\S^\s=DD$, with its new bosonic worldvolume coordinate.
This is also implied by the identity $(fú)_{DDD\d}$: without $\ff$'s
\begin{equation}
f_{D[D}{}^P ú_{D]P\d} = 0
\end{equation}
or explicitly
\begin{equation}
©^a_{Œ[º|}©_{a|©]¶} = 0
\end{equation}
which is violated.  This also requires the introduction of $\s$:
\begin{equation}
\label{E:smore}
f_{D[D}{}^P ú_{D]P\d} = ú_{DD\s}\ff_{D\d}{}^\s
\end{equation}
This identity is essentially the same as the ``dual" identity $(fú)_D^{¯¯\varpi}$
\begin{equation}
\label{E:doowop}
f_{DP}{}^{[¯}ú^{¯]P\varpi} = ú^{¯¯\varsigma}\ff_{D\varsigma}{}^\varpi
\end{equation}
which instead requires the introduction of $\varsigma$,
under the identifications
\begin{equation}
f_{DP}{}^¯ = ú_{DP\d}¼,âú^{¯P\varpi} = f_{DD}{}^P¼,âú^{¯¯\varsigma} = ú_{DD\s}¼,â\ff_{D\varsigma}{}^\varpi = \ff_{D\d}{}^\s
\end{equation}

The case $(fú)_{DPP\varpi}$ yields the $©$-matrix relation (\ref{E:gammma}), as the dual to (\ref{E:Gamma}),
\begin{equation}
\label{E:mama}
f_{D(P}{}^¯ ú_{P)¯\varpi} = \ff_{D\varpi}{}^{\un\p} ú_{PP\un\p}
\end{equation}
Note that this requires $\ff_{D\varpi}{}^ ±0$.


\subsection{Gauss}

The analysis of the previous subsections has helped to elucidate the details of spacetime and the worldvolume, and their coordinate invariances, through the cataloging of the metric $ú$ and structure constants $f,\ff$.  A complete analysis would require the factorization of the $U$ matrices (in terms of the $ú$'s just found) onto the CGW coefficients $c$ (\ref{E:CGW}), thus determining the Gauss gauge transformation.  Here we discuss a few general features, and leave a detailed analysis for a future paper.

The classification of $\U$ constraints ($c$'s) strongly resembles that of $\S$ constraints ($ú$'s, which are also CGW coefficients), as well as that of the $\dd$'s themselves:
\begin{equation*}
\begin{array}{lllll}
\dd_A & \D^\A & \S^\A & \U_{\vec A} & \\
\hline
S & \sâ & PS + DD & \vec S & P\s + \un\p S + D\d \\
D & \d & PD + ¯S & \vec D & P\d + \un\p D + ¯\s + \varpi S \\
P & \un\p & PP + ¯ D+êS & \vec P & P\un\p + ¯\d + \varpi D +ê\s + \varsigma S \\
¯ & \varpi & P¯ + ê D & \vec ¯ & P\varpi + \un\p ¯  + ê\d +\varsigma D \\
ê & \varsigma & Pê + ¯¯ & \vec ê & P\varsigma + \un\p ê + ¯\varpi
\end{array}
\end{equation*}

For example, for the old $¯$, we would find
\begin{equation}
U^{D\p}_{D\p} = U^{Œ\b}_{º\a} = ¶^Œ_º ¶_\a^\b + ú^{¯Œ\b} ú_{¯º\a}
= ¶^Œ_º ¶_\a^\b - (¶^Œ_© ¶^\b_\c)(¶_º^© ¶_\a^\c) = 0
\end{equation}
and thus no corresponding fermionic gauge parameter nor field.
On the other hand, for the new one no such cancelation is possible.
Furthermore, the representation for $\vec D$, a mixed symmetry trispinor, can be seen already by noting that it includes terms from both $P\d$ and $\un\p D$.  

The existence of $\vec D ¾ \un\p D + ...$ is why, unlike the usual version of superstring theory, the first-class constraints close without the introduction of a dimension-3 operator $D\p D$.  Its Lagrange multiplier is a ``partner" to $Î$, and will allow a Lagrangian formulation of F-theory that is manifestly symmetric with respect to both supersymmetry and the generalized Lorentz symmetry.



\subsection{Summary}

Now the nontrivial invariant tensors are
\begin{equation*}
\begin{array}{llll}
\hfil f_{AB}{}^C \hfil & \hfil \ff_{A\B}{}^\C \hfil & \hfil ú_{AB\C} / ú^{AB\C} \hfil \\
\hline
 & & ú_{PP\un\p} = ú_{ab\un\c} \\
 & & ú^{PP\un\p} = ú^{ab\un\c} \\
 & \ff_{P\un\p}{}^\s = \ff_{a\un\b}{}^\s & \\
 & \ff_{P\varsigma}{}^{\un\p} = \ff_a{}^{\s\un\b} & \\
f_{DP}{}^¯ = ©_{bŒ©} & & ú_{DP\d} = ©_{bŒ©} \\
f_{DD}{}^P = ©^c_{Œº} & & ú^{P¯\varpi} = ©^a_{º©} \\
 & \ff_{P\varpi}{}^\d = ©_a^{º©} & ú_{P¯\varpi} = ©_a^{º©} \\
 & & ú^{DP\d} = ©^{bŒ©} & \\
 & \ff_{D\un\p}{}^\d = -ý_{\un\b Œ}{}^© & ú_{D¯\un\p} = ý_{\un\c Œ}{}^º \\
 & \ff_{D\varpi}{}^{\un\p} = -ý_Œ^{\un\c º} & ú^{D¯\un\p} = ý^{\un\c Œ}_º \\
 & & ú_{DD\s} = ú^{¯¯\varsigma} = ú_{Œº\s} \\
 & \ff_{D\d}{}^\s = \ff_{D\varsigma}{}^\varpi = \ff_{μ}{}^\s & \\
 & \ff_{¯\varpi}{}^\s = \ff_{¯\varsigma}{}^\d = ú^{Œº\s} & ú^{DD\s} = ú_{¯¯\varsigma} = ú^{Œº\s} 
\end{array}
\end{equation*}
(where $ú_{Œº\s}=-ú_{ºŒ\s}$, etc.).
There are also many tensors involving $S$, which are easily determined by Lorentz-group theory, and their ``dual" tensors involving $ê$.

Thus the flat-space current algebra now looks like this:
\begin{subequations}
\begin{align}
iÓD_Œ,D_ºÕ & = ©^a_{Œº}P_a ¶ - 2ú_{Œº\s}\s ¶ \\
i[D_Œ,P_a] & = ©_{aŒº}¯^º ¶ - 2©_{aŒº} \d^º ¶ \\
i[P_a,P_b] & = f_{ab}{}^ê ê ¶ - 2ú_{ab\c}\p^\c ¶ \\
iÓD_Œ,¯^ºÕ & = ©_{SŒ}{}^º ê - 2ý_{\a Œ}{}^º \p^\a ¶ \\
i[P_a,¯^Œ] & = -2©_a^{Œº}\varpi_º ¶ \\
iÓ¯^Œ,¯^ºÕ & = -2ú^{Œº\s}\varsigma ¶
\end{align}
\end{subequations}
and similar relations with $S$ or $ê$ on the left-hand side.

Also, we have for infinitesimal global symmetry transformations
\begin{equation}
¶X^A = ñ^A - üñ^B X^C f_{BC}{}^A + ...¼,⶧_\A = - ñ^B §_\C \ff_{B\A}{}^\C
\end{equation}
and in particular for supersymmetry
\begin{equation}
¶Î^Œ = ·^Œ¼,â¶X^a = ü·^º Î^© ©_{º©}^a¼;â¶\vartheta_Œ = 0¼,⶧_{\un\a} = ·^º\vartheta_© ý_{\un\a º}{}^©
\end{equation}
These leave invariant the derivatives (ignoring Lorentz coordinates)
\begin{equation}
d_Œ = »_Œ +üÏ^º ©^a_{ºŒ} »_a + ...¼,â\d^Œ = »^Œ - Î^º ý_{\un\a º}{}^Œ »^{\un\a} + ...
\end{equation}
where $»_Œ=»/»Ï^Œ$, $»_a=»/»x^a$, $»^Œ=»/»\vartheta_Œ$, and $»^{\un\a}=»/»§_{\un\a}$.


\section{Conclusions}

We have shown that consistent backgrounds for the supersymmetric case require the worldvolume be described by a superspace that includes fermionic worldvolume coordinates for $û$-symmetry, or more generally for a set of first-class constraints.  We also found that selfduality could be generalized to the entire affine Lie algebra, explaining the origin of the constraints that eliminate doubled fermionic coordinates for spacetime.  

These results open up many new avenues of exploration:

\bitem{$\diamond$} Massless backgrounds need to be examined: 
For example, the naive orthogonality of the vielbeins needs to be reduced to the usual bosonic one by the torsion constraints (as in T-theory the large OSp symmetry is broken to O(D,D)).

\bitem{$\diamond$} The minimal 3D case should be examined in detail, for comparison to results for (linearized) supergravity obtained previously by methods not directly related to string theory.  The appearance of superconformal transformations in the linearized 3D theory suggests the relevance of the usual supergroups, which might clarify the algebraic structure.

\bitem{$\diamond$} In general, and especially for the critical dimension D = 10, the ``internal" $Y(§)$ coordinates and their duals need to be included.  This may require some modification from our earlier treatment.  In particular, addition of new currents implies new $\U$ constraints (from $\widetilde Y$), but a similar extension to worldvolume coordinates ($\D$ and $\S$) is also expected.  The tensor hierarchy might naturally show up here.

\bitem{$\diamond$} We have yet to describe generalization to higher D (before including $Y$).

\bitem{$\diamond$} In this paper we have used mostly the Hamiltonian formalism; translation to the Lagrangian formalism would have several important uses.  For example, the new choice of $¯$ allows the covariantization of supersymmetry with respect to the ``Lorentz" symmetry of the Lagrangian formalism.  This is a direct consequence of the appearance of a fermionic gauge field $\Theta_\tau$, the ``other part'' of the Lorentz spinor $Î$, which also allows a nontrivial supersymmetry transformation of the bosonic gauge field.  This will require a further analysis of the $\U$ constraints.

\bitem{$\diamond$} A related issue is the appearance of $ $ as a worldvolume coordinate in the Hamiltonian formalism.  This needs to be eliminated, or identified with or (at least) related to the usual ``time" $ $, possibly in a way similar to how the second $ $ was removed by worldvolume sectioning for D = 4 \cite{Linch:2015qva}.

\bitem{$\diamond$} The significance of $\vartheta$ as a worldvolume coordinate, and its corresponding superspace,  is yet to be understood:  Why should it not appear also for the superstring and even for the superparticle?  In particular, it would eliminate the objectionable weight-3 $D©D'$ appearing in the (otherwise weight-2) first-class super Virasoro algebra of superstring theory \cite{Siegel:1985xj}, as $D'$ would then be a piece of a Gauss $\U$ constraint (if T-theory can be obtained from F-theory by applying only the bosonic $\U$ constraint).

\bitem{$\diamond$} It might also be interesting to investigate any relationships to the ``superembedding" approach \cite{Bandos:1995zw, Howe:1996mx}:  The main constraint there is, in our notation,
\begin {equation}
(\d^ΠX)^a = 0
\end{equation}
which in our formalism can be recognized as part of the selfduality condition (\ref{E:SD2}) with the imposition of the usual mixed first and second-class constraints
\begin {equation}
D_Π= 0
\end{equation}

\section*{Acknowledgements}
W{\sc dl}3 is supported by National Science Foundation grants PHY-1521099 and PHY-1620742. 
W{\sc s} is supported by NSF grant PHY-1620628.


\appendix
\newpage

\section{Representations}

The covering groups listed in (\ref{E:table}) allow the use of spinor notation to simplify algebra.
Bosons tend to be bispinors:
We sometimes use the notation $(Œº)$, $[Œº]$, $ӌºÕ$, $ҌºÔ$ to indicate matrices that are symmetric, antisymmetric, SO-traceless (symmetric), and Sp-traceless (antisymmetric).

The representations that appear in D=1-6 are
\begin{equation*}
\label{E:tableq}
\hskip-5pt
\begin{array}{rccccccc}
\rm D & 1 & 2 & 3 & 4 & 5 & 6 \\
\rm H_D & \rm GL(1) & \rm GL(2) & \rm Sp(4) & \rm Sp(4; {\bf C} ) & \rm USp(4,4) & \rm SU*(8) \\
\hbox{spinor} & 1¢1 & 2¢2 & 4 & 4¢Ð4 & (8,2) & (8,2,1)¢(8',1,2) \\
_{\un Œ} & _à & _Œ,_{Ќ} & _Œ &  _{Œ,ÀŒ} & _{ŒŒ'} & _{ŒŒ'},^Œ{}_{Ќ'} \\
\hbox{st vector} & 1¢1¢1 & 3¢3 & 10 & 16 & 27 & 28¢28' \\
_a & _{à,0} & _{(Œº),(Ќк)} & _{(Œº)} & _{ŒÀº} & _{ҌºÔ} & _{[Œº]},^{[Œº]} \\
\hbox{wv vector} & 1¢1 & 3¢1 & 5¢1 & 5¢Ð5¢1¢1 & 27¢1 & 63¢1(¢70) \\
_{\un\a} & _à & _{ŒÐº} & _{[Œº]} & _{[Œº],[ÀŒÀº]} & _{[Œº]} & _Œ{}^º(,^{[Œº©¶]})
\end{array}
\end{equation*}

For example, for D = 3, according to the Sp(4) symmetry of the Hamiltonian formalism, we have indices $Œ$ = {\bf 4} for the spinors, $\a$ = $ҌºÔ$ (Sp-traceless antisymmetric) = {\bf 5} for worldvolume $§$(not $ $)-vectors, and $a$ = $[\a\b]$ = $(Œº)$ = {\bf 10} for the spacetime vectors.
For D = 5 the spinors are pseudoreal, so there is a USp(2) R-symmetry even for the ``minimal" case.
For D = 6 the maximal case is required, with USp(2)$^2$.
(It will be important for the Fierz identity in the following section.)

Here is a list of some of the spacetime $©$-matrices, worldvolume $ý$-matrices, and $ú_{ab\un\c}$ relating the two.
The latter two will be derived from the first in the following subsection, but we list them here for convenience.
For D = 3 we have
\begin{subequations}
\begin{align}
©_a^{©¶} = ©_{Œº}^{©¶} & = ¶_{(Œ}^© ¶_{º)}^¶ \\
©_{a©¶} = ©_{Œº,©¶} & = -C_{©(Œ}C_{º)¶}
\end{align}
\end{subequations}
\vskip-.33in
\begin{subequations}
\begin{align}
ý_©^{\un\a ¶} = ý_©^{Œº,¶} & = ¶_©^{[Œ}C^{º]¶} \\
ý_{\un\a ©}{}^¶ = ý_{Œº,©}{}^¶ & = ¶^¶_{[Œ}C_{º]©}
\end{align}
\end{subequations}
\begin{equation}
ú_{ab\un\c} = ú_{Œº,©¶,·½} = C_{[·|(Œ}C_{º)(©}C_{¶)|½]}
\end{equation}
where $C$ is the antisymmetric, hermitian (imaginary) Sp(4) metric,
with normalization
\begin{equation}
C^{Œ©}C_{º©} = ¶^Œ_º
\end{equation}

For D = 4,
\begin{subequations}
\begin{align}
©_a^{©À¶} = ©_{ŒÀº}^{©À¶} & = ¶_Œ^© ¶_{Àº}^{À¶} \\
©_{a©À¶} = ©_{ŒÀº,©À¶} & = C_{Œ©}C_{ÀºÀ¶}
\end{align}
\end{subequations}
\begin{equation}
ú_{ab\un\c} = (ú_{ŒÀº,©À¶,·½},ú_{ŒÀº,©À¶,À·À½}) = (C_{ÀºÀ¶}C_{Œ[·}C_{½]©},C_{Œ©}C_{Àº[À·}C_{À½]À¶})
\end{equation}
The expressions for $ý$ are the same as in D = 3, but also the complex conjugate equations.

For D = 5,
\begin{subequations}
\begin{align}
©_a^{©¶} = ©_{Œº}^{©¶} & = ¶_{Ҍ}^© ¶_{ºÔ}^¶ \\
©_{a©¶} = ©_{Œº,©¶} & = -C_{©ÒŒ}C_{ºÔ¶}
\end{align}
\end{subequations}
\begin{equation}
ú_{ab\un\c} = ú_{Œº,©¶,·½} = C_{[·|Ҍ}C_{ºÔÒ©}C_{¶Ô|½]}
\end{equation}
The expression for $ý$ is again the same as D = 3.

We stop at D = 6, where $_a=(_{[μ]},^{[μ]})$:
\begin{subequations}
\begin{align}
©_a^{©¶} = (©_{Œº}^{©¶},©^{Œº,©¶}) & = (¶_{[Œ}^© ¶_{º]}^¶,0) \\
©_{a©¶} = (©_{Œº,©¶},©^{Œº}_{©¶}) & = (0,¶_{[©}^Œ ¶_{¶]}^º)
\end{align}
\end{subequations}
\begin{equation}
(ý_Œ{}^º){}_©{}^¶ = ¶_Œ^¶ ¶_©^º
\end{equation}
\begin{equation}
ú_{Œº}{}^{©¶}{}_·{}^½ = ¶_{[Œ}^½ ¶_{º]}^{[©} ¶_·^{¶]}
\end{equation}
For D = 5 and 6 we have ignored the R-symmetry factors that accompany the $©$'s; these will be discussed below where necessary.

We also have the convention for contracting bispinor versions of vector (or other) indices, whenever there is a (anti)symmetry on the two spinor indices
\begin{equation}
\label{E:half}
VÉW ­ üV^{Œº}W_{Œº}
\end{equation}
to avoid double counting (but no $ü$ when expressed in terms of vector indices).  For example, for D = 3
\begin{equation}
¶_a^b = ¶_a^c ¶_c^b = ü(¶_{(Œ}^· ¶_{º)}^½)(¶_{(·}^© ¶_{½)}^¶) = ¶_{(Œ}^© ¶_{º)}^¶
\end{equation}


\section{Matrix identities}


\subsection{Dirac matrix identity}
\label{A:gama}

In this appendix we examine (\ref{E:gama}) 
\begin{align*}
©^{(a}_{Œ©}©^{b)©º} & = - ú^{ab\un\c}ý_{\un\c Œ}{}^ºâ(ý_{ Œ}{}^º ¾ ¶_Œ^º)\\
©_{(a|Œ©}©_{|b)}^{©º} & = - ú_{ab\un\c}ý_Œ^{\un\c º}â(ý_Œ^{ º} ¾ ¶_Œ^º)
\end{align*}
in various D, determining the extension $\p£\un\p$.

The case D = 1 is rather trivial.  (We leave it as an exercise for the reader.)
D = 2 requires including extra scalars $Y$ even for the ``minimal" case, so will be left for a later paper.

For D = 3 (but see (\ref{E:half})),
\begin{subequations}
\begin{align}
©_{(a}^{·ú}©_{b)½ú} & = ©_{Œº}^{·ú}©_{©¶,½ú} + (Œº ª ©¶) \\
& = (¶_{(Œ}^· ¶_{º)}^ú)(-C_{½(©}C_{¶)ú}) + (Œº ª ©¶) \\
& = - C_{½(©}C_{¶)(Œ}¶_{º)}^· + (Œº ª ©¶) \\
& = -ü(C_{[ú|(Œ}C_{º)(©}C_{¶)|Ï]})(¶_½^{[ú}C^{Ï]·}) \\
& = - üú_{Œº,©¶,úÏ}ý_½^{úÏ,·} = - ú_{ab\un\c}ý_½^{\un\c ·}
\end{align}
\end{subequations}
Note that in this case
\begin{equation}
\un\a = [Œº] = (ҌºÔ, ) = (\a, ) = 5 ¢ 1
\end{equation}
for the traceless and trace pieces.

The case D = 4 is similar, but actually a bit easier because of the separation of undotted ({\bf 4}) and dotted ({\bf Ð4}) indices.  
We then have
\begin{subequations}
\begin{align}
©_{(a}^{·Àú}©_{b)½Àú} & = (¶_Œ^· ¶_{Àº}^{Àú})(C_{©½}C_{À¶Àú}) + (ŒÀº ª ©À¶) \\
& = - ü(C_{ÀºÀ¶}C_{Œ[ú}C_{Ï]©})(¶_½^{[ú}C^{Ï]·}) \\
& = - üú_{ŒÀº,©À¶,úÏ}ý_½^{úÏ,·}
\end{align}
\end{subequations}
and the complex conjugate equations ($ŒªÀŒ$).  So there is also the complex conjugate part of $ý$ above, since
\begin{equation}
\un\a = [Œº]¢[ÀŒÀº] = (5 ¢ 1)¢(Ð5 ¢ 1)
\end{equation}
(There are now two ``$ $"'s, which is resolved for D = 4 by worldvolume sectioning \cite{Linch:2015qva}.)

D = 5 is more similar to D = 3, but trading symmetrization for traceless antisymmetrization (ignoring R-symmetry USp(2) indices, which factorize trivially for this identity):
\begin{subequations}
\begin{align}
©_{(a}^{·ú}©_{b)½ú} & = (¶_{Ҍ}^· ¶_{ºÔ}^ú)(-C_{½Ò©}C_{¶Ôú}) + (Œº ª ©¶) \\
& = -ü(C_{[ú|Ҍ}C_{ºÔÒ©}C_{¶Ô|Ï]})(¶_½^{[ú}C^{Ï]·}) \\
& = - üú_{Œº,©¶,úÏ}ý_½^{úÏ,·}
\end{align}
\end{subequations}
so now we have
\begin{equation}
\un\a = [Œº] = (ҌºÔ, ) = (\a, ) = 27 ¢ 1
\end{equation}

D = 6 simplifies due to lack of a Lorentz Sp metric, but has doubling for 
$_{\un Œ}=(_{ŒŒ'},^Œ{}_{Ќ'})$ (but the R-symmetry USp(2)'s are unimportant again) and $_a=(_{[Œº]},^{[Œº]})$:
\begin{subequations}
\begin{align}
©_{(a}^{·ú}©_{b)½ú} & = (©_{Œº}^{·ú}©^{©¶}_{½ú}+0,0+©_{©¶}^{·ú}©^{Œº}_{½ú}) \\
& = ([¶_{[Œ}^· ¶_{º]}^ú][¶_½^{[©} ¶_ú^{¶]}], Œº ª ©¶) \\
& = -([¶_{[Œ}^Ï ¶_{º]}^{[©} ¶_ú^{¶]}][¶_Ï^· ¶_½^ú], Œº ª ©¶) \\
& = - (ú_{Œº}{}^{©¶}{}_ú{}^Ï ý_Ï{}^ú{}_½{}^·,Œº ª ©¶)
\end{align}
\end{subequations}
so now we have
\begin{equation}
\un\a = _Œ{}^º = (_Œ{}^º - \hbox{tr}, ) = (\a, ) = 63 ¢ 1
\end{equation}
There is also a {\bf 70} that doesn't appear in the (flat-space) supersymmetry algebra.


\subsection{Lorentz identity}
\label{A:Lorentz}

We now show how the constraint (\ref{E:Lorentz}) 
\begin{equation*}
©_a^{Œ©}©^b_{Œº} - ©_{aŒº}©^{bŒ©} = f_{Sa}{}^b ©^{S©}_º
\end{equation*}
requires the current ``$S$", and determines the corresponding ``Lorentz" group (H$_{\rm D}$).  
(The constraint (\ref{E:duwop}) determines $ê$ in the ``dual" way.)
The cases D = 3,4,5 are all similar, because the group for each case is symplectic.
For D = 3,
\begin{equation}
(©_{Œº})^{½ú}(©^{©¶})_{·ú} - (©^{©¶})^{½ú}(©_{Œº})_{·ú} = ¶_{(º}^{(¶|}(¶_{Œ)}^½ ¶_·^{|©)} + C^{½|©)}C_{Œ)·})
\end{equation}
which tells us
\begin{equation}
(©_Œ{}^º)_©{}^¶ = ¶_Œ^¶ ¶_©^º + C^{¶º}C_{Œ©}
\end{equation}
\begin{equation}
(f_Œ{}^º)_{©·}{}^{¶½} = ¶_{(©}^{(¶|}(¶_{Œ)}^½ ¶_·^{|º)} + C^{½|º)}C_{Œ)·})
\end{equation}
$(©^S)_Œ{}^º$ easily can be identified as group generators for the symplectic group in the defining representation by raising and lowering indices with the metric $C$:
\begin{equation}
C^{·©}C_{º½}(©_Œ{}^º)_©{}^¶ = - ¶_{(Œ}^· ¶_{½)}^¶
\end{equation}
(I.e., they are then a basis for all symmetric matrices.)  Reality conditions then identify the group as Sp(4).
(Reality conditions on F-spinors for each D are the same as those on ordinary spinors in D dimensions, as those on supersymmetry generators.  Only the size of the spinors has been doubled.)

D = 4 is again simpler:  We find
\begin{equation}
(©_{ŒÀº})^{½Àú}(©^{©À¶})_{·Àú} - (©^{©À¶})^{½Àú}(©_{ŒÀº})_{·Àú} = ¶_{Àº}^{À¶}(¶_Œ^½ ¶_·^© + C^{½©}C_{Œ·})
\end{equation}
We can thus again identify
\begin{equation}
(©_Œ{}^º)_©{}^¶ = ¶_Œ^¶ ¶_©^º + C^{¶º}C_{Œ©}
\end{equation}
and now
\begin{equation}
(f_Œ{}^º)_{©À·}{}^{¶À½} = ¶_{À·}^{À½}(¶_Œ^¶ ¶_©^º + C^{¶º}C_{Œ©})
\end{equation}
(and the complex conjugate equations).  Complexification then gives the group Sp(4,{\bf C}).

D = 5 is again similar to D = 3:
\begin{equation}
(©_{Œº})^{½ú}(©^{©¶})_{·ú} - (©^{©¶})^{½ú}(©_{Œº})_{·ú} = ¶_{Òº}^{Ò¶|}(¶_{ŒÔ}^½ ¶_·^{|©Ô} + C^{½|©Ô}C_{ŒÔ·})
\end{equation}
\begin{equation}
(©_Œ{}^º)_©{}^¶ = ¶_Œ^¶ ¶_©^º + C^{¶º}C_{Œ©}
\end{equation}
\begin{equation}
(f_Œ{}^º)_{©·}{}^{¶½} = ¶_{Òº}^{Ò¶|}(¶_{ŒÔ}^½ ¶_·^{|©Ô} + C^{½|©Ô}C_{ŒÔ·})
\end{equation}
Pseudoreality gives the group USp(4,4).

D = 6 is again simplest:  The 2 $©©$ terms contribute separately to different matrix elements, due to lack of an Sp metric $C$.  (Thus each $©$ must have mixed up and down spinor indices.)  For example,
\begin{equation}
(©_{Œº})^{½ú}(©^{©¶})_{·ú} - (©^{©¶})^{½ú}(©_{Œº})_{·ú} = ¶_{[º}^{[¶|}(¶_{Œ]}^½ ¶_·^{|©]} + 0)
\end{equation}
\begin{equation}
(©_Œ{}^º)_©{}^¶ = ¶_Œ^¶ ¶_©^º
\end{equation}
\begin{equation}
(f_Œ{}^º)_{©·}{}^{¶½} = ¶_{[º}^{[¶|}(¶_{Œ]}^½ ¶_·^{|©]})
\end{equation}
Pseudoreality now gives the group U*(8) = SU*(8)$°$GL(1), since this $©^S$ is a basis for arbitrary matrices.
(The extra real scale GL(1) is related to the USp(2)$^2$ R-symmetry and corresponding $Y$ coordinates, which we have mostly ignored in this paper.)


\subsection{More identities}
\label{A:more}

As we have seen, all the $fú$ identities go pretty much the same way, so we'll give just a few more examples, and restrict to just D = 4.  
First we look at (\ref{E:ddual})
\begin{equation*}
©^a_{ú(·} ý^\b_{½)}{}^ú = ú^{ah\b}©_{h·½}
\end{equation*}
or in D=4
\begin{equation}
©^{ŒÀº}_{úÀ·} ý^{©¶,ú}_½ ­ (¶_{À·}^{Àº}¶_Œ^ú)(¶_½^{[©}C^{¶]ú})
= ú^{ŒÀº,úÀÏ,©¶}©_{úÀÏ,½À·} ­ (C^{ÀºÀÏ}C^{Œ[©}C^{¶]ú})(C_{ú½}C_{ÀÏÀ·})
\end{equation}
(and the complex conjugate equation).  This is satisfied, as both sides equal
\begin{equation}
¶_{À·}^{Àº}¶_½^{[©}C^{¶]Œ}
\end{equation}
(The proof of (\ref{E:dddual}) is almost identical.)

A similar example is (\ref{E:moreexplicitly}).  It also involves $©ý$, but with opposite symmetrization:
\begin{equation*}
©_{aú[·}ý^{\b ú}_{½]} = \ff_{a\varsigma}{}^\b ú_{·½}^\varsigma
\end{equation*}
We now have instead for the left-hand side
\begin{equation}
©_{ŒÀº,úÀ·}ý_½^{©¶,ú} = C_{ÀºÀ·}¶_½^{[©}¶_Œ^{¶]}
\end{equation}
so identifying $\varsigma=úÀÏ$, we find
\begin{equation}
ú^\varsigma_{À·½} = - ú^\varsigma_{½À·} = -¶_½^ú ¶_{À·}^{ÀÏ}
\end{equation}
\begin{equation}
\ff_{ŒÀº,úÀÏ}{}^{©¶} = C_{ÀºÀ¶}¶_Œ^{[©}¶_ú^{¶]}
\end{equation}
It thus determines part of $\varsigma$, but leaves out $\varsigma=[Œº]¢[ÀŒÀº]$.

Similar remarks apply to (\ref{E:whatever}),
\begin{equation}
©_{c·[Œ|}ý_{\ff |º]}{}^· = ú_{Œº\s}\ff_{c\ff}{}^\s
\end{equation}
which it strongly resembles.  The result is essentially the same, with some raising and lowering of indices, and it applies to $\s$ rather than its dual $\varsigma$.

We'll give a complete determination of $\s$ (and thus $\varsigma$) in the following subsection.


\subsection{Fierz identity}
\label{A:Fierz}

Before looking at the Fierz identity, we examine (\ref{E:smore}) and (\ref{E:doowop}), which are similar to it, but have opposite symmetrization:
\begin{equation}
©^a_{Œ[º|}©_{a|©]¶} = ú_{º©\s}\ff_{Œ¶}{}^\s = ú_{º©}{}^\varsigma \ff_{Œ\varsigma ¶}
\end{equation}
We again look only at D = 4, but there are 3 cases to consider:
\begin{subequations}
\begin{align}
©^a_{ŒÀº}©_{a©À¶} & = C_{Œ©}C_{ÀºÀ¶} = ú_{Àº©\s}\ff_{ŒÀ¶}{}^\s \\
©^a_{ŒÀº}©_{aÀ©¶} & = C_{Œ¶}C_{ÀºÀ©} = ú_{ÀºÀ©\s}\ff_{Œ¶}{}^\s
\end{align}
\end{subequations}
and the complex conjugate of the latter.  From these we find
\begin{subequations}
\begin{align}
ú_{©Àº,·À½} = -C_{©·}C_{ÀºÀ½}¼,&â\ff_{ŒÀ¶}{}^{·À½} = ¶_Œ^· ¶_{À¶}^{À½} \\
ú_{ÀºÀ©\s} = C_{ÀºÀ©}¼,&â\ff_{Œ¶}{}^\s = C_{Œ¶}
\end{align}
\end{subequations}
so that $\s$ = {\bf 16} + {\bf 1} + {\bf 1}.
In D = 3 a similar analysis finds $\un\s$ = {\bf 5}$¢${\bf 1}:  Two representations are needed to reduce to left and right-handed constraints in T-theory.
In this case we then see that $\un\s\un\p\un\varsigma$ are the same Sp(4) representations, just as $SPê$ ({\bf 10}).
Similar results are found for D = 5 ($\s$ = {\bf 27}$¢${\bf 1}) and D = 6 ($\s$ = {\bf 63}$¢${\bf 1}).

We now look at the Fierz identity (\ref{E:Fierz}):
\begin{equation*}
©^a_{(Œº}©^b_{©)¶}ú_{ab\un\c} = 0
\end{equation*}
This can be satisfied only in D = 3,4,6,10:
It requires worldvolume scalars $Y$ (and their duals $\widetilde Y$) to complete the total dimension to those values.
In this paper we do not consider these scalars nor D = 10, so we'll look at only the cases D = 3,4,6.
For convenience, and without loss of generality, we can consider the form
\begin{equation}
©^a_{(Œº}©^b_{©¶)}ú_{ab\un\c} = 0
\end{equation}

It's also convenient to note that, e.g.¼for D = 3, with $a=Œº$ and $b=©¶$,
\begin{equation}
¶_a^b = ©_a^{©¶} = ©^b_{Œº} = ¶_{(Œ}^© ¶_{º)}^¶
\end{equation}
using the notation of the previous section (with obvious generalization for $©^b_{Œº}$).  Then
\begin{equation}
©^a_{Œº}©^b_{©¶}ú_{ab\un\c} = ú_{Œº,©¶,·½} = C_{[·|(Œ}C_{º)(©}C_{¶)|½]}
\end{equation}
so symmetrization in $Œº©¶$ vanishes.

Similarly for D = 4
\begin{equation}
©^a_{ŒÀº}©^b_{©À¶}ú_{ab,·½} = ú_{ŒÀº,©À¶,·½} = C_{ÀºÀ¶}C_{Œ[·}C_{½]©}
\end{equation}
and the same holds.  ($©$ vanishes if indices aren't mixed.)

For D = 6 the spinor indices are really $_{\un Œ}=(_{ŒŒ'},^Œ{}_{Ќ'})$.  
We now need to take the USp(2) R-symmetry indices $Œ'$ and $Ќ'$ into account, by introducing an extra factor of these USp metrics into the definition of the $©$'s.
This has the effect of making them symmetric matrices under the interchange of such pairs of indices, allowing $P_{[Œº]}$ and $P^{[Œº]}$ to come out of $ÓD,DÕ$:
\begin{equation}
ÓD_{ŒŒ'},D_{ºº'}Õ ¾ C_{Œ'º'}P_{[Œº]}¼,âÓD^Œ{}_{Ќ'},D^º{}_{к'}Õ ¾ C_{Ќ'к'}P^{[Œº]}
\end{equation}
Furthermore, both chiralities of spinor indices are necessary to get the dual $¯$ out of $[D,P]$:
\begin{equation}
[D_{ŒŒ'},P^{[º©]}] ¾ C_{Œ'º'}¶_Œ^{[º}¯^{©]º'}¼,â[D^Œ{}_{Ќ'},P_{[º©]}] ¾ C_{Ќ'к'}¶^Œ_{[º}¯_{©]}{}^{к'}
\end{equation}
(This differs from S- and T-theory, where the smaller SU*(4) Lorentz symmetry allows the tensor $·_{Œº©¶}$.)
As a result, scalars $Y^{Œ'к'}$ are again required for the currents $\mho$  \cite{Linch:2015fca} that appear in 
\begin{equation}
ÓD_{ŒŒ'},D^º{}_{к'}Õ ¾ \mho_Œ{}^º{}_{Œ'к'}
\end{equation}
(The index $Œ'к'$ = {\bf (2,2)} of USp(2)$^2$, giving the extra 4 ``internal" dimensions for the critical $6+4 = 10$.)
Details will be left for a future paper.

{
\small
\linespread{1.1}\selectfont
\raggedright


} 


\end{document}